\documentstyle[12pt]{article}
\begin{document}
\noindent
{\Large\bf UNIVERSE, STRUCTURE OF THE}\par
\bigskip
\noindent
Prof. Dr. Hans J. Haubold, UN Outer Space Office, Vienna
International Centre, P.O. Box 500, A-1400 Vienna, Austria\par
\bigskip
\noindent
Prof. Dr. A.M. Mathai, Department of Mathematics and
Statistics, McGill University, 805 Sherbrooke Street West,
Montreal, Quebec, Canada H3A 2K6\par
\vspace{4cm}
\noindent
{\large\bf ABSTRACT}\par
\medskip
\noindent
At present the most plausible theory of the origin of the
universe is that it formed from the explosion of an
extremely hot and dense fireball several billion years ago.
During the
first few seconds after the Big Bang, the energy density was
so great that only fundamental particles (leptons, quarks,
gauge bosons) existed. As the universe cooled and expanded
after the Big Bang, nuclei and atoms formed and condensed into
galaxies and stars and systems of them. The fundamental
particles and a wide range of gravitational aggregates of them
constitute the small-scale and large-scale structure of the
present universe. The current knowledge of the elements of the
structure of the universe based on the standard Big Bang model
and the standard model of fundamental particles is
considered.
\clearpage
\noindent
{\bf TABLE OF CONTENTS}\par
\bigskip
\noindent
{\bf INTRODUCTION}\par

\bigskip
\noindent
\begin{tabbing}
1.1.1.1. \= \=\kill \\
1. \>STRUCTURE OF THE ANCIENT UNIVERSE\\
1.1\> Prehistoric Astronomy\\
1.2\> Egypt\\
1.3 \> Mesopotamia\\
1.4 \>India and Persia\\
1.5 \>The Puranas and the Brahmapaksa\\[1cm]

2. \>MACROCOSMOS: FROM ANAXIMANDER TO EINSTEIN\\
2.1 \>Greek and Roman World\\
2.2\> Aristotle's Universe\\
2.3 \>Copernicus' Universe\\
2.4\> Newton's Universe\\
2.5\> Einstein's Universe\\
2.6 \>Big Bang \\
2.7\> Beyond the Big Bang \\[1cm]

3.\> MICROCOSMOS: FROM LEUKIPPUS TO YUKAWA\\
3.1\> Atomism\\
3.2\> Atom\\
3.3\> Electron\\
3.4\> Atomic Nucleus and Photon\\
3.5 \>Neutron\\
3.6\> Fundamental Particles\\[1cm]

4.\> WEAK, ELECTROMAGNETIC, AND STRONG INTERACTIONS:\\ 
 \>  STRUCTURE OF THE MICROCOSMOS\\
4.1 \>Fundamental Interactions\\
4.1.1\>Gravitational interaction\\
4.1.2\>Electromagnetic interaction\\
4.1.3\>Weak interaction\\
4.1.4\>Strong interaction\\
4.2\> Classification of Fundamental Particles\\
4.3\> How Fundamental Particles Interact\\
4.3.1\>Graviton\\
4.3.2\>Photon\\
4.3.3\>Intermediate bosons\\
4.3.4\>Gluons\\[1cm]

5.\> GRAVITATIONAL INTERACTION:\\
  \> STRUCTURE OF THE MACROCOSMOS\\
5.1 \>Structural Elements of the Universe\\
5.1.1\> Planck scales\\
5.1.2\> Universe\\
5.1.3\> Galaxies\\
5.1.4\> Stars\\
5.1.5\> White dwarfs and neutron stars\\
5.1.6\> Black holes\\
5.2\> Evolution of the Universe\\
5.3\> Large-scale Distribution of Matter\\
5.4\> Morphology of Galaxies\\
5.5.\> Quasars\\
5.6.\> Stellar Evolution\\[1cm]

6.\> MATHEMATICS AND PHYSICS AND THE STRUCTURE \\
\>OF THE UNIVERSE\\
6.1\> General Relativity and Quantum Field Theory\\
6.2\> Macroscopic Dimension of Space\\       
6.3\>Microscopic Dimension of Space\\[1cm]
   
\> Glossary\\[1cm]

\>   List of Works Cited\\[1cm]
   
\>   Further Reading List\\

\end{tabbing}
\clearpage

\noindent
{\large\bf INTRODUCTION}\par
\medskip
\noindent
Every culture, from the ancient tale-spinners of the Indus
Valley to the modern technocrats of the Silicon Valley, has
held its own unique view of the cosmos. Astronomy, the oldest
science mainly based on observation , cannot be separated from
physical law and mathematics. Big advances in
space science and technology in this century have allowed
astronomers to look both deeper into space and farther back in
time, thereby discovering a close connection between
particle physics and cosmology. The study of the creation,
evolution, and structure of the universe has become a
legitimate subject for astronomers, physicists, and
mathematicians.\par 
\smallskip

The article presents an outline of the current knowledge about
the structure, more precisely the structural elements, of the
universe from the point of view of the standard Big Bang model
(dubbed macrocosmos) and the standard model of fundamental
particles (dubbed microcosmos). In this regard an important
point has to be made that the article does not consider the
time dependent phenomena that ought to characterize the evolution
of the universe but is focused on the structural elements
of the universe as discovered by observation at the present
stage of the evolution of the universe. One can ask the
question of how it is possible to understand the universe on
the basis of its structural elements? The ancient atomic
philosophers may have answered: This objection is similar to
that of someone saying, how can the entire poem of Homer, the
Iliad, be composed of only 24 letters, those 24 letters of the
Greek alphabet. Hidden in that thought is, that the richness
of the poem is not compromised by the fact, that a method exists
to
ultimately trace them back to a few, very elementary and
simple elements, such as the formal contents of the Greek
language to the 24 characters, that are being referred to as
the Greek alphabet. However, the degree of the current
understanding of the structural elements of the universe
(compared with the 24 letters of the Greek alphabet) reveals
the fact, that the scientific establishment is still far away
from understanding the beauty of the structure and evolution
of the universe as a whole (compared with the Iliad).\par
\smallskip
The article is divided into six parts. The first three parts
recount the development of macrophysical and microphysical
knowledge from prehistory to the present. Part four is devoted
to the elements of the standard model of fundamental
interactions, namely the leptons, quarks, and gauge bosons.
The fifth part discusses the structural elements of the
universe, from the large-scale structure down to the stars.
Finally, the sixth part emphasizes the importance of
mathematics for the physics and astronomy of the structure of
the universe.
\clearpage

\section{STRUCTURE OF THE ANCIENT UNIVERSE}
\subsection{Prehistoric Astronomy}

From half a millennium before 4000 BC, long barrows -
elongated burial mounds of earth - were arranged so as to be
aligned on the rising and setting of bright stars. 
In the middle of the third millennium this technique was
transferred to circular grave architecture, the so-called round
barrows, and in this form it lasted for well over a thousand
years. Why this particular phase in prehistorical astronomical
activity is so important is that it testifies to the
marriage of astronomy with geometry. There is clear evidence
that
the Sun and the Moon were observed. Stonehenge, shown in
Fig.1, is the most famous prehistoric megalith (standing-stone
monument) in Europe and lies north of Salisbury, England.
Stonehenge had an exceptionally long history of use as a
religous center and is also believed to have functioned as an
astronomical observatory. \par
\begin{center}
FIGURE 1.
\end{center}
\subsection {Egypt}

Some concern with astronomy had already been shown in a
cosmology associated with the rulers Seti I (r.c.1318-c.1304
BC) and
Ramses IV (r.c.1166-c.1160 BC). The Egyptians had by then long
been
adept at measuring time and designing calendars, using simple
astronomical techniques. They too aligned their buildings on
the
heavens. Some early Egyptian sources speak of a cult relating
the
Sun-god Re (represented in art with a man's body and a falcon's
head surmounted by a solar disk) and an earlier creator-god
Atum. The cult of Re-Atum
was well established by the time of the first great pyramids,
that is, about 2800 BC. At first this was centered mainly on a
temple to the north of the old Egyptian capital of Memphis. The
place was known by the Greeks as Heliopolis, 'City of the Sun',
but by the Egyptians as On (located northeast of Cairo, Egypt).
By historical times, the priests of
Heliopolis had laid down a cosmogony that held Re-Atum to have
generated himself out of Nun, the primordial ocean. His
off-spring were the gods of air and moisture, and only after
them, and as their off-spring, were Geb, the Earth god, and
Nut,
the sky goddess, created. The deities of Heliopolis, (the
Great Ennead) were made up with Osiris, Seth, Isis, and
Nephthys,
the off-spring of Geb and Nut.

\subsection{Mesopotamia}

In the dynasty of Hammurabi (r.c.1792-c.1750 BC) in Babylonia
not all gods can be identified with
stars. The three highest - Anu, Enlil, and Ea - corresponded
to the heavens, Earth, and water.

\subsection{Indian and Persian Astronomy}

The oldest of the Vedic writings in Hinduism, the Rigveda
(dating from between 1500 and 500 BC), gives more than
one account of the creation of the world (Fig.2). The main
version\par 
\begin{center}
FIGURE 2.
\end{center}
\noindent
is that the world was made by the gods, as a building of wood,
with
heaven and Earth somehow supported by posts. Later it is
suggested that the world was created from the body of a
primeval
giant. This last idea gave rise to the principle, found in
later
Vedic literature, that the world is inhabited by a world soul.
Various other cosmogonies followed, with the creation of the
ocean sometimes being given precedence, and place being made
for the creation of the Sun and the Moon. There is a certain
circularity in it all, however, since heaven and Earth are
generally regarded as the parents of the gods in general; and
water was sometimes introduced into the parentage. The Vedic
literature gives no clear indication that mathematical
techniques
for describing the motions of heavenly bodies were discussed in
India before the 5th century BC.

\subsection{The Puranas and the Brahmapaksa}

Although influenced by texts going back to Vedic times, and by
Iranian sources too, the Puranas are Sanskrit writings about
primordial times with cosmological
sections dating from the 4th to 16th  centuries AD of the
Christian era. The Puranas deal with five topics: the creation of
the Universe; the destruction and re-creation of the Universe,
including the history of humankind; the genealogy of the gods;
the reigns of the Manur; and the history of the lunar and solar
dynasties.
The Earth is now represented as a flat circular disc, with a
world-mountain, Meru, at its center. The Meru is anchored to
the yasti, which symbolizes the axis of the universe. The
mountain is surrounded by
alternating rings of sea and land, so that there are seven
continents and seven seas. Wheels are conceived to carry the
celestial bodies, these turning around the star at the north
pole by Brahma, using cords made of mind. This cosmology was
taken over by Jainism, a monastic religion which like Buddhism
denies the authority of the Veda, but from the 5th and 6th
centuries onwards it was undermined by the influx of a new form
of Greek cosmology, with pre-Ptolemaic roots. In short,
Aristotelianism reached India (see North, 1995).
\section{MACROCOSMOS: FROM ANAXIMANDER TO\\
EINSTEIN}

\subsection{Greek and the Roman World}

Anaximander of Miletus (c.610-545 BC) is said to have made a
map of the inhabited world,
and to have invented a cosmology that could explain the
physical
state of the Earth and its inhabitants. The infinite universe
was said to be the source of an infinity of worlds, of which
ours
was but one, that separated off and gathered its parts together
by their rotary motion. Masses of fire and air were supportedly
sent outward and became the stars. The Earth was some sort of
floating circular disc, and the Sun and the Moon were
ring-shaped
bodies, surrounded by air. The Sun acted on water to produce
animate beings, and people were descended from fish.\par
\smallskip
Anaximenes (c.545 BC) elaborates on Anaximander's ideas, and
argues that
air is the primeval infinite substance, from which bodies are
produced by condensation and rarification, he produces logical
arguments based on every day experience. Again he introduces
rotary motions as the key to understanding how the heavenly
bodies may be formed out of air and water.\par
\smallskip
Pythagoras of Samos (c.560-c.480 BC) took the cosmic ideas of
Anaximander and\\ 
Anaximenes one stage further, saying that the universe was
produced by
the heaven inhaling the infinite so as to form groups of
numbers.
The Pythagoreans proposed a geometrical model of the universe,
involving a central fire around which the celestial bodies
move in circles. The central fire was not the Sun, although the
Earth was certainly of the character of a planet to it. To
account for lunar eclipses, the Pythagoreans postulated a
counter
Earth.\par
\smallskip
The discovery that the Earth is a sphere was traditionally
assigned to Parmenides of Elea (c.515-c.450 BC), who was also
said to have discovered that the Moon is
illuminated by the Sun. A generation later, Empedocles
(c.484-c.424 BC) and
Anaxagoras (c.500 BC) seem to have given a correct qualitative
account of
the reason for solar eclipses, namely the obscuration of the
Sun's face by the intervening Moon.\par 
\smallskip
The discovery of the sphericity of the Earth, and of the
advantages of describing the heavens as spherical, captured the
imagination of the Greeks of the time of Plato (c.428-c.347 BC)
and Aristotle (384-322 BC), in
the 4th century, and of one man particular: Eudoxus of Cnidus
(c. 400-c.355 BC), who produced a very remarkable planetary
theory based entirely on spherical motions.

\subsection {Aristotle's Universe}

At the time of the Greek philosopher and naturalist Aristotle
(384-322 BC), the Earth and the universe were seen as
constructed out of five basic elements: earth, water, air,
fire,
and ether. The natural place of the motionless Earth was at the
centre of that universe. The stars in the heavens were made up
of
an
indestructible substance called ether (aether) and were
considered
as
eternal and unchanging. Aristotle's cosmology was the first
`steady
state' universe. The other basic elements -
water, air, fire - were earthly elements. The celestial bodies,
including the Sun, the planets, and the stars, were considered
to
be attached to rigid, crystalline spheres, which were supposed
to
revolve in perfect
circles about the Earth. The three innermost spheres, closest
to
the
Earth, contained water, air, and fire, respectively.
The Egyptian-Greek astronomer Ptolemy (c.100-170 AD) left
the Earth to be located at the
centre of the universe but ascribed to the Sun and the planets
a
new place within the cosmological views of his time. The Sun
and
the
planets would revolve in small circles whose centers revolve in
large
circles about the Earth (``epicycles''). The element of
perfection
and beauty
ascribed to the divine heavens remained the circle out of
which the orbits of heavenly bodies were composed. Basically,
Ptolemaic ideas were devised to accomodate astronomical
observations of planetary motions. His
``Almagest'' described the motions of the heavens on this
geocentric basis,
employing wheels within wheels (epicycles) to obtain a
plausible match with
observations. Ptolemy's geocentric system dominated western
thought
on astronomy (including cosmology) until the time of
Copernicus,
fourteen centuries later (Fig.3).\par 
\begin{center}
FIGURE 3.
\end{center}

Eratosthenes (c.270-c.190 BC) was the first to accurately
measure the
radius of the Earth in 196 BC by determining the minimum angle
between the Sun's direction and the vertical at Alexandria on
the day of the summer solstice. He knew that a zero angle
occurs
approximately when the Sun was at its highest point at the city
of
Syene (now
Aswan), and he  knew the base of the triangle, i.e. the
distance
from Alexandria to Syene.

\subsection{Copernicus' Universe}

Copernicus (1473-1543), a Polish astronomer, placed the Sun
at the center of the
solar system with the Earth orbiting around the Sun, thus
proposing
a heliocentric cosmology (De revolutionibus orbium
coelestium, 1543).
This entirely
new basis of cosmological consideration did not fit the
observations much better than the Aristotelian-Ptolemaic system
but
was justified by the ``divine'' principle of simplicity in
comparison to the
rather complicated construction using epicycles as employed in
Ptolemy's
cosmology. However, like Aristotle and Ptolemy, Copernicus
retained the conventional idea that the planets moved in
perfectly
circular orbits and continued to believe that the stars were
fixed
and unchanging.
Kepler (1571-1630), a German astronomer and physicist, adopted
the
Copernician system but introduced the concept of  planets with
elliptical
orbits. He still believed
that the Sun was the centre of the universe. 
The cosmologies of Aristotle and Ptolemy had nevertheless been
abandoned.

\subsection{Newtons' Universe}

The Italian philosopher Bruno (1548-1600) laid the
groundwork for Newtonian cosmology by emphasizing that the
universe
is
infinite and stars are scattered outward through infinite space
(see, however, Fig.4).
Bruno even went so far as to say that stars are Suns, perhaps
with
\begin{center}
FIGURE 4.
\end{center}
\noindent
orbiting planets and life on them (``De l'infinito,
universo e mondi'', 1584). This far-reaching statement
signaled a transfer of attention away from the planets in the
solar
system to
the stars in the Milky Way. The laws describing the
behavior
of planets are the same laws which describe the behavior of the
stars. That the
physical laws are of universal nature, and can be applied on
Earth
as well as in the heavens, was discussed on a philosophical
basis by the French philosopher and mathematician Descartes
(1596-1650). He compared the universe with a
giant clock, obeying mechanical laws which later on had a major
influence on Newton's (1643-1727) thinking. It was his belief
that
the
universe was infinite and that the primary qualities of the
universe were mathematical in nature.\par 
A major step in observational
techniques was achieved through the invention of the reflecting
telescope between, say, 1545 and 1559 by the Britain Leonhard
Digges (c.1520-1559). But it was not until 1609 that the
Italian astronomer and
physicist Galileo (1564-1642) realized observationally
that the Milky Way is actually a collection of
individual
stars. Galileo also observed
mountains on the Moon and discovered four satellites around
Jupiter. 
He had taken the first step toward deducing the
structure of the Milky Way. The discovery of heavenly bodies
that
evidently did
not circle the Earth, and his support for the Copernician
heliocentric cosmology were described in his work
``Dialogo sopra i
due massimi sistemi del mondo, Tolemaico e Copernicano''
(1632).\par
\smallskip
Cassini (1625-1712), an Italian-French astronomer, used the
telescope to make accurate
measurements of the ``dimensions of the universe'',
particularly
determining the distance of the Earth from the planet Mars. A
rigorous mathematical foundation of Descartes' notion of the
universe as a giant mechanical clock was provided by Newton's
theory of gravity and his laws of motion (``Philosophiae
naturalis principia mathematica'', 1687). From these he
explained
Galileo's results on falling bodies, Kepler's three laws of
planetary motion and the motion of the Moon, Earth and tides.
Newton clearly realized that gravity is the dominant force for
understanding the structure of the universe; however,
he argued that the universe must be static in a famous
letter he sent to the theologian Richard Bentley (1662-1742) in
1692.
Mainly for religious reasons, at this time, constancy and
stability
were associated with the perfection of God and change with
friction and
decay. Philosophical and religious ideas served as the
scaffolding
upon which scientific systems of thought developed. It was
Bentley
who derived for the first time, based on Newton's gravitational
theory, what is still considered to be one of the fundamental
constants of nature, the gravitational constant.\par 
In
Kant's (1724-1804) cosmology the gravitational
attraction of stars for each other was exactly balanced by the
orbital motions of the stars and he argued that forces can act
at
a distance without the necessity for a transmitting medium. In
1755
Kant proposed the nebular hypothesis for the formation of the
solar
system. In 1788 Laplace (1749-1827) attempted a mathematical
proof of the stability of the  solar system
(Syst\`{e}me du monde, 1796). Stability
and order of the universe were considered as eternal principles
in
the heavens and on Earth. The cosmology of Copernicus, as
refined
by
Kepler, is now believed to be essentially correct.

\subsection{Einstein's Universe}

In 1915 Einstein (1879-1955) put forth his general relativity
theory (a new theory of gravity);  he applied this theory to
cosmology in 1917. The theory describes
gravity as a distortion of the geometry of space and time.
Unlike
Newton's theory of gravity, general relativity was consistent
with
special relativity, which Einstein had introduced in 1905.
Cosmology, based on general relativity, broadened the problem
into
one of finding a model of the space-time structure of the
universe.
Einstein's original solutions of
his gravitational field equations left the universe in a stable
state of static equilibrium and he provided physical conditions
required to maintain such static (time-independent) equilibrium
(the ``cosmological constant''). Only in 1922 Friedmann
(1888-1925)
succeeded in finding solutions
of
Einstein's field equations that evolved in time describing an
expanding (or contracting, if one cares to reverse the sense of
time) universe.\par 
\smallskip
Persuasive observational
evidence that the universe is indeed expanding and changing in
time
was found by Hubble (1889-1953) in 1929 while employing the
technique known as Doppler shift for measuring the red shift
of
colors in the spectrum of nebulae. ``Modern cosmology'' may be
said
to have began with Einstein (1917) and Friedmann (1922, 1924),
based
on observations of cosmological relevance made by Slipher 
(1875-1969) in 1914
and Hubble in 1929. 
Slipher discovered the redshift of nebulae, later on identified
by
Hubble to be entire galaxies similar to the Milky Way;
however, until 1929 their cosmological significance remained
obscure. In 1929
Hubble
had counted a great number of galaxies (to determine their
distribution
throughout the observable universe), and plotted the galaxie's
redshifts against magnitudes for the brightest E-type cluster
galaxies (Hubble diagram). Hubble found evidence that the
outward
speed of a galaxy is directly proportional to its distance away
from the observer (Hubble law). This observational fact was
exactly
what would
be expected if the universe is expanding, as discussed in a
paper
of
Lema\^{\i}tre (1894-1966) in 1927. Hubble's diagram reveals a
linear
increase of the magnitude of galaxies with increasing redshift.
In
1956, Hoyle and Sandage developed the $q_0$ criterion which
could
be used to distinguish one cosmological model from others.
Lema\^{\i}tre's model and Hoyle's `steady state' model were
ruled out by
predicting $q_0=-1$, whereas Hubble's linearity gave $q_0=+1$.
Cosmology focused on the search for two numbers: $H_0$, the
rate
of expansion of the universe at the position of the Milky Way;
and $q_0$, the deceleration parameter, characterizing
the
change of the rate of universal expansion. The value of $q_0$
is
believed to lie in the range between 0 and 0.5; the value of
$H_0$
is thought to be uncertain by a factor of about 2
($H_0=100hkm\; sec^{-1} Mpc^{-1}, \mbox{where}\; 0.5\leq h \leq
1$).
Theoretical ideas on which Big Bang cosmology
is
based were  contained in the publications
of Friedmann and Lema\^{\i}tre by 1930 (for an in-depth review
see North, 1965).

\subsection{Big Bang} 

Astrophysical arguments were introduced into the cosmological
models with
Gamow's (1946) prediction that helium (and possibly heavier
elements)
were generated
at an early stage of the evolution of the universe. At this
time it
was believed that the relative abundance of cosmic nuclei
represents truly cosmic abundances. Based on Gamow's arguments
the
cosmological criterion of the origin of chemical elements in
the
primeval fireball was substantiated  by Alpher and Herman
(1949, 1950) and Alpher et al. (1953). However, after Burbidge
et
al. (1957) it became evident that the bulk of the chemical
elements
beyond helium where not synthesized at the early stages of the
expansion of the
universe but in the stars. Only the synthesis of helium (Hoyle
and
Tayler, 1964), which is not produced in sufficient quantities
in
stars, and deuterium (Peebles, 1966), which is destroyed during
galactic evolution, continued to need Gamow's
primordial nucleosynthesis to arrive at reasonable abundances
as
observed in the universe. Later on Wagoner et al. (1967) were
able
to show that in addition to $D$, $^3He$, and $^4He$, the only
other
cosmologically significant element was $^7Li$. In 1965 the
microwave background radiation of 3 degrees Kelvin was
discovered
by Penzias and Wilson (1965), as predicted by Alpher and Herman
in
1949 based on Gamow's considerations of a hot and dense origin
of
the universe.\par 
\smallskip
Today the Big Bang cosmological model, the
temperature of the microwave background radiation, the
abundances
of $D$, $^3He$, $^4He$, $^7Li$, and the astrophysically
observed
average density of galactic material are starting points for
developments in cosmology.

\subsection{Beyond the Big Bang}

In the Big Bang cosmology, the universe has been expanding
throughout its history. Mathematical calculations would
suggest that the temperature and the density were infinite at
the
instant of the Big Bang. The universe is supposed to originate
in
the Big Bang, but the mathematical and physical structure of
the
model does not permit matter to originate. Additionally, the
Big
Bang model leaves unanswered several important questions
regarding
(i) the number of protons and neutrons in the universe,
relative to
the number of photons, (ii) the large-scale homogeneity of the
observed universe, (iii) the actual density of the universe
which
is
close to the critical density, and (iv) the origin of density
perturbations from which small-scale (stars and systems of
them)
and large-scale (galaxies and systems of them)
inhomogeneities have been developed (Kolb and Turner,  
1990).\par
\smallskip
A new approach to explaining some of the questions left over by
Big
Bang cosmology began with Guth's (1981) paper inventing an
``inflationary'' period in the evolution of the universe. The
phase transition associated with the break-up of the unified
force in the Grand Unification epoch could leave the universe in
a state of false vacuum, in which the vacuum has a very high
energy density. This vacuum energy density acts like a cosmic
repulsion and the universe embarks on an exponential expansion
which ``inflates'' the universe by a factor of $10^{30}$ in a
brief instant of time. This inflationary period ends when the
vacuum energy density transforms into matter and radiation and
the expansion of the universe continues. Besides
offering explanation of the four questions above, it
makes one concrete prediction: The present universal expansion
should exhibit flatness to the extent of
$\Omega_0=1\pm\epsilon$
with $\epsilon<10^{-6}$, if inflation was sufficient to extend
to
a distant $\geq cH_0^{-1}$ today. Observations of the luminous
matter
content of the universe reveal only $\Omega_0<0.002$, and the
quantity of dark or non-baryonic material necessary to explain
the
flat rotation curves of spiral galaxies and the virial
equilibrium
of large groups and clusters of galaxies requires at most
$\Omega_0\approx 0.2$. The only way that $\Omega_0=1$ from
inflationary
cosmological models can be reconciled with observation is by
the
existence of non-baryonic weakly interacting particles. The
unknown
nature of dark matter led to considerable activities in
fundamental
particle physics to speculate about new basic constituents of
matter. Additionally, the inflationary universe picture has
stimulated far-reaching speculation about the ultimate origin
of the universe. \par
\smallskip
Cosmology based on general relativity has thus moved close to
fundamental particle physics based on quantum field theory. A
new
challenge emerged for physical theory and observation subsumed
under the term `quantum gravity'. Interaction between
mathematics and physics, as exemplified by the role of
Riemannian 
geometry in general relativity and functional analysis
in quantum mechanics, became lively again. Cosmology is
currently
based on the two fundamental theories in twentieth-century
physics,
general relativity and quantum field theory. General relativity
describes the gravitational force on an astronomical scale,
while
quantum field theory describes the weak, electromagnetic, and
strong
interaction of fundamental particles. A formal quantization of
general relativity leads to unphysical infinities; encompassing
gravity in quantum field theories leads to a new connection of
theoretical physics to modern mathematics, called string
theory.
However, there is still no quantum cosmology. Particularly, an
inflationary era may have taken place in the early universe,
but
there is no proof that it did so (Manin, 1981; Schmid, 1992). 

\section{MICROCOSMOS: FROM LEUKIPPUS TO\\
YUKAWA}

\subsection{Atomism}

With regard to the structure of the universe, the belief
persists, that certain objects are fundamental and that others
are derived, in the sense that the latter are composed of the
former.  In one version of this distinction, the fundamental
objects are particles, or points of matter. Although the view
of which objects qualify as fundamental particles has changed
many times, the notion that the universe is ultimately made of
such material points, moving through space, has endured in 
some form ever since the theory of atomism was first proposed
by the Greeks Leucippus (5th century BC) and Democritus
(c.460-c.370 BC)
 in the 5th century BC (Whittaker, 1951 and 1953/1989).\par
\smallskip
An atom is the smallest unit of matter that is recognizable as
a chemical element.  Atoms of different elements may also
combine into systems called molecules, which are the smallest
units of chemical compounds (Fig.5).  In all these processes,
atoms may be considered as the ancient Greeks imagined them to
be: The ultimate building blocks of matter.  When stronger
forces are applied to atoms, however, the atoms may break up
into smaller parts.  Thus atoms are actually composites and
not units, and have a complex inner structure of their own.\par

\begin{center}
FIGURE 5.
\end{center}

The first recorded speculations that matter consisted of atoms
are found in the works of Leucippus and Democritus.  The
essence of their views is that all phenomena are to be
understood in terms of the motions, through empty space, of a
large number of tiny and indivisible bodies. The name ``atom''
comes from the Greek word ``atomos'', for ``indivisible''.
According to Democritus, these bodies differ from one another
in shape and size, and the observed variety of substances
derives from these differences in the atoms composing them.\par
\smallskip
Greek atomic theory was not an attempt to account for specific
details of physical phenomena.  It was instead a philosophical
response to the question of how change can occur in nature.
Little effort was made to make atomic theory quantitative -
that is, to develop it as a physical theory for the
study of matter.  Greek atomism, however, did introduce the
valuable concept that the nature of everyday things was to be
understood in terms of an invisible substructure of objects
with unfamiliar properties.  Democritus stated this especially
clearly in one of the few sayings of his that has been
preserved:  ``color exists by convention, sweet by convention,
bitter by convention, in reality nothing exists but atoms and
the void.''\par
\smallskip
Although adopted and extended by such later ancient thinkers
as Epicurus (341-270 BC) and Lucretius (c.95-55 BC), Greek
atomic theory had strong
competition from other views of the nature of matter. One such
view was the four-element theory of Empedocles. These
alternative views, championed by Aristotle among others, were
also motivated more by a desire to answer philosophical
questions than by a wish to explain scientific phenomena. 

\subsection{Atom}

The atomic theory languished until the 18th and early 19th
centuries, when physicists and chemists revived it to explain
the properties of gases and some of the facts of chemistry. In
these theories the fundamental particles, the atoms, remained
indivisible points.  The discovery in the late 19th and the
20th centuries that atoms were composite, rather than
indivisible, set the stage for modern discoveries about
fundamental particles (Pais, 1986; Whittaker, 1951 and
1953/1989).\par
\smallskip
When interest in science revived in Europe in the 16th and
17th centuries, enough was known about Greek atomism to form
the basis for further thought.  Among those who revived the
atomic theory were Pierre Gassendi (1592-1655), Robert Boyle
(1627-1691), and
especially Isaac Newton.  The latter part of Newton's book
``Optics'' is a series of detailed speculations on the atomic
nature of matter and light, indicating how some of matter's
properties are to be understood in terms of atoms.\par
\smallskip
In the 19th century, two independent lines of reasoning 
strengthened the belief in the atomic theory.  Both approaches
also began to reveal some quantitative properties of atoms. 
One approach, pioneered by John Dalton (1766-1844), involved
chemical
phenomena. The other, involving the behavior of gases, was
carried out by physicists such as Rudolf Clausius (1822-1888)
and James
Clerk Maxwell (1831-1879).\par
\smallskip
Dalton's main step forward was his introduction of atomic
weights.  Dalton studied the elements then known and analyzed
the data of their reactions with one another.  He discovered
the law of multiple proportions, which states that when
several distinct reactions take place among the same elements,
the quantities that enter the reactions are always in the
proportions of simple integers - that is, 1 to 1, 2 to 1, 2 to
3, and so on.  From this came the concept that such reacting
quantities contain equal numbers of atoms and are therefore
proportional to the masses of individual atoms. Dalton gave
the lightest known element, hydrogen, an atomic weight of 1,
and developed comparative atomic weights for the other known 
elements accordingly.\par
\smallskip
The study of gases in terms of atomic theory was begun by 
Daniel Bernoulli (1700-1782) in the 18th century.  Bernoulli
showed 
that the pressure exerted by a gas came about as the 
result of collisions of the atoms of the gas with the 
walls of its container.  In 1811, Amedeo Avogadro (1776-1856) 
suggested that equal volumes of different gases, under the 
same conditions of pressure and temperature, contain equal 
numbers of atoms. Avogadro himself never estimated the
magnitude of this value, although it is now known as the
Avogadro number.

\subsection{Electron}

The history of  particle physics has gone through four
stages. In the first stage, Joseph J. Thomson (1856-1940)
discovered
(1897), by studying electricity passing through gases, that
all atoms contain certain particles, called electrons, that
carry a negative electric charge. Because atoms are
electrically neutral, there must be balancing positive charges
somewhere in the atom.  Ernest Rutherford (1871-1937) proposed
(1911),
based on a series of experiments by Hans Geiger (1882-1945)
and Ernest Marsden (1889-1970) that these positive charges are
concentrated in a
very small volume, called the atomic nucleus, at the center of
the atom.\par
\smallskip
By the end of the 19th century almost all scientists had
become convinced of the truth of the atomic theory.  By that
time, however, evidence was just beginning to accumulate that
atoms are not in fact the indivisible particles suggested by
their name. One source of such evidence came from studies
using gas discharge tubes. In such tubes, a gas at low
pressure is subjected to intense electrical forces. Under
these conditions, various colored glows are observed to
traverse the tube. One blue glow at one end of the tube,
around the electrode known as the cathode, was observed for a
wide variety of gases.  The glow was shown by Joseph Thomson in
1897 to involve a stream of negatively charged
particles with a charge-to-mass ratio, indicating the existence
of a particle with a very small mass on the atomic scale. These
particles were called electrons, and they were
soon recognized to be a constituent of all atoms. That is, 
atoms are not indivisible but contain parts.\par
\smallskip
In the late 19th and the early 20th century it was also found
that some kinds of atoms are not stable.  Instead they
transform spontaneously into other kinds of atoms. For
example, uranium atoms slowly change into lighter thorium
atoms, which themselves change into still lighter atoms,
eventually ending up as stable atoms of lead. These
transformations, first observed by Antoine Henri Becquerel
(1852-1908),
came to be known as radioactivity, because the atomic changes
were accompanied by the emission of several types of
radiation.\par
\smallskip
Atoms are ordinarily electrically neutral. Therefore the
negative charge of the electrons in an atom must be balanced
by a corresponding positive charge.  Because the electrons
have so little mass, the positive constituents of an atom must
also carry most of the atom's mass. The obvious question arose
as to how these varied parts are arranged within an atom. The
question was answered in 1911 through the work of Ernest
Rutherford and his collaborators. In their experiments they
passed alpha particles - a type of radiation emitted in some
radioactive decays - through thin gold foils. They observed
that in some instances the alpha particles emerged in the
opposite direction from their initial path.  This suggested a 
collision with a heavy object within the atoms of the gold.
Because electrons are not massive enough to produce such large
deflections, the positive charges must be involved. Analyzing
the data, Rutherford showed that the positive charge in an
atom must be concentrated in a very small volume with a radius
less than $10^{-12}$ cm, or one ten-thousandth the size of
the whole atom.  This part of the atom was called the
nucleus.

\subsection{Atomic Nucleus and Photon}

Rutherford proposed an atomic model in which the atom was held
together by electrical attraction between the nucleus and the
electrons.  In this model the electrons traveled in relatively
distant orbits around the nucleus. The model eventually proved
successful in explaining most of the phenomena of chemistry
and everyday physics.  Subsequent studies of the atom divided
into investigations of the electronic parts of the atom, which
came to be known as atomic physics, and investigations of the
nucleus itself, which came to be known as nuclear physics.
This division was natural, because of the immense difference
in size between the nucleus and the electron orbits and the 
much greater energy needed to produce nuclear as compared to
electronic changes.\par
\smallskip
The Rutherford model of the atom, however, had to face two 
immediate problems.  One was to account for the fact that 
different atoms of the same element behaved in physically and
chemically similar ways.  According to the Rutherford model,
electrons could move in any of the infinite number of orbits
allowed by Newtonian physics. If that were so, different atoms
of the same element could behave quite differently. This is
actually a problem for any atomic model based on Newtonian
physics, and it had already been recognized by Maxwell in
1870. The other problem was that, according to the principles
of electromagnetism, electrons should continuously emit
radiation as they orbit in an atom. This would cause the
electrons to lose energy and to spiral into the nucleus.\par
\smallskip
An important step toward solving these problems was taken by
Niels Bohr (1885-1962) in 1913.  According to Bohr, the
electrons in atoms
cannot exist in arbitrary orbits. Instead they are found only
in certain ``states''.  The states in which they can exist are
those in which the angular momentum of their orbits is an
integer multiple of $h/2\pi$, where ``$h$'' is a quantity known
as
Planck's constant. This constant had been introduced by Max
Planck (1858-1947) in his theory describing blackbody
radiation.\par
\smallskip
According to the Bohr model of the atom, there is a so-called
ground state for any atom. This ground state has the lowest
energy allowed to the atom, and it is the same for all atoms
containing the same number of electrons. An atom normally
exists in this ground state, which determined the observed
properties of a given element. Furthermore, according to Bohr,
no radiation is emitted by an atom in its ground state. This
is because energy must be conserved in the radiation process,
and no available state of lower energy exists for the atom to 
balance any energy lost through radiation.\par
\smallskip
An atom can be removed from its ground state only when enough
energy is given to it, by radiation or collisions, to raise an
electron to an ``excited'' state. When the atom is
excited, it will usually emit electromagnetic radiation
rapidly and return to the ground state. The radiation is
emitted in the form of individual packets or quanta, of light,
called photons. Each photon has an energy equal to the
difference between the energy of the excited states and the
ground state of the atom. According to a formula developed by
Planck and Einstein, this energy corresponds to a
specific wavelength of the emitted light. Using this assumption
about the allowed angular momenta for electrons, Bohr was able
to calculate the precise wavelengths in the spectrum of the
simplest atom, hydrogen. \par
\smallskip
In 1869, Dimitri I. Mendel\'eev (1834-1907) stated the rule that
chemical elements arranged according to the value of their atomic
weights exhibit a clear periodicity of properties. Eventually,   
 Bohr was able to extend his atomic theory to describe,
qualitatively, the chemical properties of all the elements.
Each electron in an atom is assigned a set of four so-called
quantum numbers. These numbers correspond to the properties
of energy, total orbital angular momentum, projection of
orbital angular momentum, and projection of spin angular
momentum. It is also assumed - as had first been suggested by
Wolfgang Pauli (1900-1958) in 1924 - that no two electrons in
an atom can
have the same values for all four quantum numbers. This came
to be known as Pauli's exclusion principle.  This principle
influences the way in which the chemical properties of an 
element depend on its atomic number, that is the number of
electrons
in each atom of the element. A maximum number of electrons
can occur for each energy level, and no more than that. For
example, the lowest energy level of an atom - the one in which
the electrons have zero orbital angular momentum - can contain
up to two electrons. The one electron in a hydrogen atom
exists at this energy level, as do the two electrons in a
helium atom.  For the next heavier atom, lithium, one of its
three electrons must exist in a higher energy state, and as a
result this electron can more easily be lost to another atom.
Those electrons with approximately the same energy are said to
form a ``shell''. \par
\smallskip
Although Bohr's model gives a qualitatively accurate 
description of atoms, it does not give quantitatively accurate
results for atoms more complex than hydrogen. In order to
describe such atoms, it is necessary to use quantum mechanics. 
This theory of atomic and subatomic phenomena was created by
Erwin
Schr\"{o}dinger (1887-1961), Werner Heisenberg (1901-1976), 
Paul Dirac (1902-1984), and Pascual Jordan (1902-1980) in the
1920s. In quantum mechanics, the electron orbits are replaced
by probability distributions that only indicate in which
regions of space each electron is most likely to be found. An
equation discovered by Schr\"{o}dinger allows this
distribution
to be calculated for each atom. From the distribution,
properties of the atom such as energy and angular momentum can
be determined. \par 
\smallskip
In the second stage, particle physics accommodated, through an
analysis of isotopes of elements, that all atomic nuclei could
be thought of as composed of two types of particles: the
proton, which carries both mass and electric charge, and the
neutron, which has about the same mass as a proton but is
electrically neutral. This model was confirmed through the
discovery (1932) of free neutrons by James Chadwick
(1891-1974).

\subsection{Neutron}

Physicists by the late 1920s were
convinced that they sufficiently understood the electronic
structure of atoms. Attention therefore turned to the nucleus.
It was already known that nuclei sometimes change into one
another through radioactive decay. Rutherford had also shown,
in 1919, that this could be accomplished artificially by
bombarding nitrogen nuclei with high-energy alpha particles. 
In the process the nitrogen nucleus is converted into an
oxygen nucleus, and a hydrogen nucleus, or proton, is ejected. 
It had further been discovered by Joseph J. Thomson, Francis W.
Aston (1877-1945), and others that for a given element the
nucleus
sometimes occurs in several different forms that differ in
mass. These chemically similar but physically distinct atoms 
were called isotopes. All of this provided evidence that
atomic nuclei also had some kind of internal structure that
could be explored through experiments and calculations.\par
\smallskip
Differences in the integer values of the electric charge and
of the mass of many nuclei soon indicated that protons were
not the only kind of particle to be found there. That is, the
electric charge of a nucleus is always exactly an integer
multiple of the charge of a proton, so knowledge of this
electric charge always indicates how many protons a nucleus
contains.  The mass of a nucleus is also approximately - but
not exactly - an integer multiple of the mass of a proton. 
For many atoms, however, these two integer values are not the
same. For example, a helium nucleus has twice the charge but
four times the mass of a proton. Clearly, nuclei contain
something other than protons.\par
\smallskip
This problem was solved in 1932 with the discovery by James
Chadwick of the neutron.  This particle has no
electric charge and is slightly more massive than a proton.
Thus most nuclei are composed of both protons and neutrons,
which collectively are known as nucleons. A helium nucleus
contains two protons and two neutrons, which correctly give
the total charge and mass of the nucleus. The isotopes of any
given element contain equal numbers of protons but different
numbers of neutrons.  For example, an isotope of hydrogen,
called deuterium, contains one proton and one neutron, and a
heavier isotope, called tritium, contains one proton and two
neutrons.\par
\smallskip
The problem then arose as to how atomic particles could be
held together in such a small region as the nucleus.  The 
force holding them had to be different from others then known
to physicists. It was stronger than the electric forces that
can break electrons away from nuclei. On the other hand, the
nuclear forces between different nuclei that are far apart are
very weak, much weaker than electric forces at such distances.

\subsection{Fundamental Particles}

The third stage of particle physics came with the 
recognition that protons, neutrons, and electrons - the 
constituents of ordinary matter - were but three of a vast 
number of similar particles, which differed only in a few 
properties, such as their mass, and in their stability against
spontaneous decay. Experiments with particle accelerators
indicated that these many subatomic particles could be readily
produced from protons and neutrons, provided that enough
energy was available to produce the additional mass of the new
particles predicted by the rules of  Einstein's
relativity theory. These discoveries in the 1940s and 1950s
indicated that the proton and neutron were not really
fundamental particles and that they would have to be
understood as part of a much larger family of similar
objects.\par
\smallskip
By 1932 nuclei were known to be composed of protons and
neutrons. It was then necessary to explain how nuclei were 
held together, and in 1935, the Japanese physicist Hideki
Yukawa
(1907-1981) predicted a smaller fundamental particle that was
the carrier
of a theorized strong interaction, one of the four fundamental
interactions, or forces. This particle, called a $\pi$-meson,
was
discovered in 1947. Since then a host of particles smaller
than protons and neutrons have been discovered in the nucleus,
all falling within two classes: Fermions, which obey the Pauli
exclusion principle, and bosons, which carry the fundamental
force. Modern nuclear physics centers on fundamental
interactions between fermions and  bosons.
Protons and neutrons are composed of particles 
representing all four forces.\par
\smallskip
In the fourth stage, modern particle physics provided a
successful explanation
for the large number of particles. 
There are six different leptons: electron (e), muon ($\mu$),
tauon ($\tau$), electron-neutrino ($\nu_e$), muon-neutrino
($\nu_\mu$), and tau-neutrino ($\nu_\tau$). There are also six
quarks denoted up (u), down (d), charm (c), strange (s), top
(t), and bottom (b). For each of these particles there exists
an anti-particle. Many of the models for particle interactions
pair the leptons and quarks into families: $(e-\nu_e), (\mu-
\nu_\mu), (\tau-\nu_\tau),$ and (u-d), (c-s), (t-b).
Experiments suggest that it is unlikely that there are more
than these families. The interactions between these fundamental
particles are mediated by gauge bosons (photon, intermediate
bosons $W^{\pm}$ and $Z^0$, gluons).

\section{WEAK, ELECTROMAGNETIC, AND STRONG\\
INTERACTIONS: STRUCTURE OF THE\\ 
MICROCOSMOS}

\subsection{Fundamental Interactions}

At sufficiently low energies, there are four types of
fundamental interactions whose
existence
is well established. Most studied are two of them: The
gravitational and the electromagnetic interactions. The
foundations of the classical (non-quantum) theory of the two
interaction types were laid long time ago by Newton and
Maxwell.\par 
\smallskip
In particle physics, the Standard Model encompasses all the
particles known today and three of the fundamental
interactions.
The basic building blocks are two sets of matter particles, the
quarks and the leptons (Table 1). These particles interact
with each other
through the exchange of
\begin{center}
TABLE 1.
\end{center}
\noindent
gauge bosons (see Section 4.3, Table 2.). The three fundamental
interactions of
the
Standard Model are the electromagnetic, the strong,
and the weak interactions, respectively. Gravity remains
outside the Standard Model,
but this does not invalidate the model as gravitational effects
on particles are far smaller than the effects of the other
interactions. The Standard Model has two components. One is the
theory
of strong interaction, called quantum chromodynamics. The other
component is the theory that gives a unified description of
electromagnetic and weak interactions, called the electroweak
theory. The physical concepts used in the Standard Model are
generalizations of concepts familiar in quantum                
 electrodynamics (Kolb and Turner, 1990; Kaku, 1993).

\subsubsection{Gravitational interaction}

Gravitational interaction that governs the motion of celestial
bodies is characterized by Newton's
constant $G = 6.7\times 10^{-8}\; g^{-1}\; cm^3\; s^{-2}$. An
excellent
approximation that describes the gravitational interaction of
two point masses $m$, a distance $r$ apart, is Newton's formula,
\begin{equation}
F = \frac{Gm^2}{r^2}.
\end{equation}
All fundamental particles are affected by gravity. The
relativistic generalization of Newton's theory of gravity
is
Einstein's theory of general relativity.

\subsubsection{Electromagnetic interaction}

Electromagnetic interaction determines the motion of charged
bodies and acts only on charged particles. In the general case,
their law of motion is described by  Maxwell's equations. In
the
quasistatic approximation, an analogue to Newton's law, the
Coulomb approximation,
\begin{equation}
F = \frac{e^2}{r^2},
\end{equation}
proves to work  well. Here, $e$ denotes the charge of
each point mass. The quantum field theory of electromagnetism,
quantum electrodynamics, is the best theory available for
describing effects of the electromagnetic force. One of this
theory's important features is its gauge symmetry, which means
that when independent changes to local field values are made at
different points in space, the equations of quantum
electrodynamics are not changed. This symmetry is ensured only
if
the quantum description of a charged particle contains an
electromagnetic field with its gauge boson, i.e., the gauge
symmetry demands the existence of the electromagnetic force and
the photon. The symmetry is also linked to the ability to
renormalize quantum electrodynamics so that it yields sensible,
finite results.\par
\smallskip 
The magnitudes of $Gm^2$ and $e^2$ depend on the choice of the
system
of units. To facilitate comparison in the framework of quantum
field theory, one combines these quantities with universal
 physical constants, the Planck constant $\hbar$ and the
velocity of
light $c$, to
obtain dimensionless constants. Thus, the nondimensional
gravitational constant (Table 3),

\begin{equation}
\alpha_g =\frac{ Gm^2}{\hbar c},
\end{equation}
\noindent
and the nondimensional electromagnetic fine structure constant
(Table 3),
\begin{equation}
\alpha_e = \frac{e^2}{\hbar c},
\end{equation}
\noindent
are obtained, $e\approx 10^{-19} C$ being the electron (proton)
charge.
There is a difference in the definition of the two constants,
$\alpha _e$ 
being in a way more universal than $\alpha_g$. The definition
of the
number $\alpha_e$ contains  fundamental constants only, whereas
the
constant $\alpha_g$ involves a mass $m$ which is, generally
speaking,
arbitrary. To eliminate this arbitrariness, it is common to fix
the value of $m$ by setting it equal to the proton mass $m_p$.
This
choice is quite natural, for the proton is one of the two
stable
particles constituting the structure of the universe; the other
one is the electron, with mass $m_e$. The choice between $m_p$
and $m_e$
is a matter of convention $(m_p \approx 1837 m_e)$.

\subsubsection{Weak interaction}

The weak interaction governs the decay of particles into
lighter
ones and acts upon all quarks and leptons, including those with
no electric charge. Historically, the first decay discovered
was
the decay of a neutron within an atomic nucleus (the
$\beta$-decay),
according to the reaction\\ 
$n\rightarrow p + e^- + \bar{\nu}\;\;(n, p, e^-,
\mbox{and}\;\;\bar{\nu}$   
denoting a neutron, a proton, an electron, and an
antineutrino,
respectively). Subsequently, the discovery of new particles was 
intensified by progress in the development of
accelerators. It turned out that all newly discovered particles
have a common property: Heavy particles decay into lighter
ones.
Numerous investigations led to the conclusion that many decays
are controlled by a unique interaction, referred to as the weak
interaction, which is characterized by the Fermi coupling
constant $g_F = 10^{-49}\; erg\; cm^3$. The corresponding
dimensionless
coupling constant for the weak interaction is (Table 3)
\begin{equation}
\alpha_w = g_F\frac{m^2 c}{\hbar^3}.
\end{equation}
\noindent
The processes of collisions of neutrinos with matter are
determined by the weak interaction as well.\par
\smallskip
There are many attempts to develop unified descriptions of all
four interactions. In quantum theory every particle is
associated with a particular field. How such a field transform
under the Lorentz transformation depends on the spin of the
particle described by the respective field. A zero-spin
particle can be described by a scalar field (Higgs field), a
spin-half particle by a spinor field, a spin-one particle by a
vector field. The Lagrangian describing the field will carry
information about the mass of the particle and its
interactions. It is possible to construct a model that
describes electromagnetic and weak interactions using a
Lagrangian which possesses invariance under two transformation
groups, U(1) and U(2).\par
\smallskip
The Higgs fields have a Lagrangian with a potential $V(\phi)$
which has non-trivial minima. If such a system comes into
contact with matter fields which are in thermal equilibrium at
some temperature $T$, then the effective potential energy will
acquire a temperature dependence. That is, $V(\phi)$ will
become $V(\phi,T)$. Such a temperature dependence can lead to
several non-trivial effects - like phase transitions - in the
early universe (Kolb and Turner, 1988). Even though the
transformation group underlying
the theory can be determined from some general principles, the
detailed transformation properties of the fields representing
specific particles cannot be derived from any fundamental
considerations. These details are fixed using the known
laboratory properties of these particles. For example, consider
the fields describing the leptons. Given a spinor field $\psi$
one can construct its `right-handed' and `left-handed'
components by the decomposition
\begin{equation}
\psi_L=\frac{1}{2}(1-
\gamma_5)\psi;\;\;\;\psi_R=\frac{1}{2}(1+\gamma_5)\psi;
\end{equation}
where $\gamma_5$ is the 4 x 4 matrix
\begin{equation}
\gamma_5=\left(\begin{array}{cc}
0 & I\\
I & 0
\end{array}
\right).
\end{equation}
In the standard electroweak theory, the right-handed components
behave as singlets (that is, they do not change) while the
left-handed components transform as a doublet (that is, under
an SU(2) transformation these fields are changed into linear
components of themselves). It is a consequence of this feature
that in the simplest electroweak theory there is no necessity
for the right-handed neutrino, $\nu_R$, and that the 
left-handed neutrino $\nu_L$ is massless. Since the
transformation
properties of the fields are put in by hand into the theory, it
is possible to generalize these models in many ways. In
particular, it is possible - though not necessary - to have
massive neutrinos in the theory. This arbitrariness is of great
importance for the existence of dark matter in the universe and
for the solar neutrino problem.

\subsubsection{Strong interaction}

The strong interaction was identified with the nuclear
interaction which acts only on quarks and is ultimately
responsible for binding protons and neutrons within the
nucleus.
The attempts to develop a consistent theory of nuclear
interaction took a long time. A breakthrough was achieved with
the progress of the dynamical theory of quark systems that led
to
the advent of quantum chromodynamics. In that scheme, the
nuclear
interaction was identified with the interaction in many-quark
systems. It is instructive to trace briefly the evolution of
the quark interpretation of nuclear interaction. To do so, we
briefly outline the quark model proposed by Murray Gell-Mann 
(b.1929) and George Zweig (b.1937) in 1964. According to this
model, 
each proton and neutron consists of three point-like particles
which are referred to as
quarks and possess a charge that is a fraction of the electron
charge $e, \pm\frac{1}{3}e$ or $\pm\frac{2}{3}e$. This
theoretical conclusion was
seemingly in contradiction to the experimental evidence that
all
the observable particles have an integer electric charge.
Nevertheless, numerous experimental confirmations of the quark
hypothesis (such as systematics of the elementary particles,
the
magnitude of the magnetic moments, the ratios of the
interaction
cross-sections, etc.) suggested that it deserves serious
consideration.  But then a profound question arose: How can the
existence of quarks be reconciled with their nonobservability
in
experiments? At present, this problem is referred to as that
of
quark-confinement. A postulate is invoked which has rather a
character of a magic: Quarks do exist, but in  bound
states. Even though no final solution of the confinement
problem is
available, one bases some expectations on the construction of a
mathematical model that claims to provide a theory of the
interaction between the quarks. It is this interaction that is
identified with the strong interaction  governing nuclear
interaction. In 1954, Chen N. Yang
(b.1922)
and Robert L. Mills (b.1927)
proposed a theory which is basically different from
electrodynamics, but accounts for the interaction caused by the
transfer of zero-mass particles. The only particle known at
that
time was the photon. The photon is the gauge particle in
 electrodynamics. That is why the Yang-Mills theory was
considered
just mathematical exercise. The picture changed, when
a
need emerged for a theory describing the dynamics of quarks. It
seemed natural to consider the massless particles introduced by
Yang and Mills to be responsible for the quark interaction.
These
particles were named gluons; by analogy with quantum
electrodynamics, the quantum field theory of electromagnetism,
one of the variants of the Yang-Mills theory is referred to as
quantum chromodynamics. While gluons are analogues of photons,
quarks are analogues of electrons. They carry not only color
charges but also ordinary electric charge. E.g., a proton
consists of three quarks $(p=uud)$, a neutron consists of three
quarks ($n=udd$), held together by continuing  exchange of
gluons. In the early 1970's the Yang-Mills equations were
subjected to more scrutiny. As a result, the constant
$\alpha_s$ was
found to exhibit quite remarkable behavior, as distinct from
quantum electrodynamics. This constant determines the
quark-quark
interaction which is currently believed to be the true strong
interaction. It should be remembered
that from the viewpoint of contemporary field theory the
interaction is mediated by gauge bosons, i.e., quanta of the
corresponding field. Energy-momentum and hence -
according to the special theory of relativity - mass is
transferred along with a quantum. Elaborate calculations have
demonstrated that the strong interaction coupling constant
$\alpha_s$
essentially depends on the energy-momentum and the mass $m$
transferred. In a way, one had encountered a mass dependence of
the constants $\alpha$ before (e.g., $\alpha_g$ and
$\alpha_w$), but quantum
chromodynamics introduces a basic difference. In this theory,
the dependence $\alpha_s (m)$ is deduced from quantum field
theory, and
not postulated, as was done earlier for the constants,
$\alpha_g$ and
$\alpha_w$, on the basis of dimensional considerations. In
addition, the
variation of the constant $\alpha_s$ with the mass $m$ has a
specific
feature: $\alpha_s$ decreases with increasing $m$. It should be
remarked
here that the terminology repeatedly used above might appear
contradictory. On the one hand, one speaks of the constants
$\alpha$; on
the other hand, one keeps stressing their dependence on $m$. In
fact,
the constants $\alpha$ are only constant at a fixed $m$; they
vary with
changing m. That is why they are referred to as
``running'' constants. The final expression for the dependence
of $\alpha_s$ on $m$
reads, in the asymptotic approximation when $m>>m_p$ (Table 3):
\begin{equation}
\alpha_s \sim\frac{a}{ln(\frac{m}{m_p})}.
\end{equation}
\noindent
The quantity $a$ depends on $N_q$, the number of the sorts of
quarks.
In a standard theory $(N_q=6), a\sim 1.$ It follows from this
formula that 
$\alpha_s\rightarrow\; 0\;\; \mbox{as}\;\; m\rightarrow
\infty$. This is the phenomenon of asymptotic
freedom. A similar dependence also follows from a more exact
expression. Unfortunately, the latter has been also obtained by
methods whose validity breaks down for $m<m_p$. A ``true''
expression
for $\alpha_s$ at small $m$ is missing, owing to the fact that
$\alpha_s$ is large,
thus rendering standard computation techniques inapplicable.
One
can only state that for a small characteristic mass $m$,
corresponding to the proton (or neutron) size, $r_N\approx
10^{-13}\;cm$, the
coupling constant is large. Furthermore, a rapid increase of
the constant $\alpha_s$, with $r$ approaching $r_N$ inhibits 
progress in solving another problem, namely, that of nuclear
forces. Today, quantum chromodynamics is considered as a theory
that describes the interactions among quarks and gluons, out of
which atomic nuclei are made.

\subsection{Classification of Fundamental Particles}

Fundamental particles are classified with respect to various
parameters. A particle classification appears to be provided by
the value of the spin, $s$. The behavior of particles depends on
whether their spin is characterized by an integer (0, 1,
2,\ldots)
or a half-integer (1/2, 3/2, 5/2,\ldots). Particles with a
half-integer
spin are referred to as fermions, while those with an integer
spin as bosons. In the framework of quantum mechanics, the
difference in the behavior of fermions and bosons is expressed
by the kind of symmetry of the wave functions describing these
particles. A system consisting of fermions obeys Pauli's
exclusion principle, as distinct from a boson system on which
no
such exclusion principle is imposed. Pauli's principle reads as
follows: No two fermions may be in exactly the same quantum
state.\par
\smallskip
An excellent illustration of Pauli's principle is the atomic
level structure underlying the periodic system of the chemical
elements.
It is known that the first period of this system is composed of
two elements, hydrogen and helium. For the first period, the
principal quantum number equals unity. The atomic states
associated with the first period are therefore determined only
by
the value of the spin projection of orbital electrons. There
are
two such values. Thus, only two elements can occur in the first
period. For the second period, the principal quantum number
equals two, giving rise to eight possible different states and
thus to eight elements, etc. The Pauli principle is one of the
foundations of the very structure of the periodic system. If
this
principle did not work, all the atomic electrons would populate
the ground energy level (i.e., the hydrogen level), and,
consequently, the periodicity of the system as well as the
valency of chemical bonding would vanish. It is the Pauli
principle that prevents atomic electrons from occupying the
energetically most favorable ground state. \par
\smallskip
Another basis for the classification of the fundamental
particles is their interaction.
All particles participating in the strong interaction are
referred to as hadrons (from the Greek word ``hadros'', meaning
``strong''). All fermions which do not participate
in
the strong interaction are called leptons. A special
place in this classification is reserved for the bosons,
particles which mediate the interactions. The hadrons, in turn,
are subdivided into the
baryons
and the mesons. The baryons are fermions; the lightest baryon
is
the proton. The hadrons with integer spin are referred to as
mesons; the lightest meson is the pion $(m_\pi \approx
140\;MeV)$.

\subsection{How Fundamental Particles Interact}

Particles interact by exchange of gauge bosons;
the exchange in the process of interaction involves not only
energy, momentum, and mass, but also the internal quantum
numbers: spin, isospin, charge, and color. The properties of
the
exchange particles in the context of quantum field theory
determine the interaction to a great extent. All exchange
particles are bosons. The properties of the exchange particles
are summarized in Table 2.\par
\begin{center}
TABLE 2.
\end{center}

\subsubsection{Graviton}

The graviton is a massless particle with zero charge and a spin
$s=\pm\;2$ and  has not been detected owing to its extremely weak
interaction. Although most
physicists
have no doubts about the existence of gravitons, some caution
is advisable, since the
quantum theory of gravitation itself is far from being complete.
Because of the weakness of the gravitational field, there is no
hope for rapid progress in detecting and investigating
gravitons. Because the graviton is a massless particle, the
gravitational interaction is long range.

\subsubsection{Photon}

Photons have spins of $s=\pm 1$, a rest mass of zero, and are
their own antiparticles. The electromagnetic interaction has a
long range because the photon is massless. 

\subsubsection{Intermediate bosons}

The weak interaction is mediated by three intermediate gauge
bosons $W^\pm$ and $Z^0$ which have masses of $\sim 80$ GeV and
$\sim 91$
GeV, respectively. Since the range of an interaction is
inversely related to the mass of the gauge boson, the weak
interaction has an extremely short range.

\subsubsection{Gluons}

The gluons, like the quarks, are not observable in the free
state. However, in the late 1970's, considerable progress in the
indirect verification of gluons was achieved by investigating
the
annihilation of high-energy positrons and electrons with hadron
generation. It turned out that three hadronic jets occur in
such
processes; two jets are attributed to quarks and the third one,
to gluons. The experimental data on the three-jet processes
accompanying positron-electron annihilation are in good
agreement
with the predictions of quantum chromodynamics, which
indirectly
confirms the existence of gluons, one of the basic elements of
that theory. The strong interaction is mediated by a set of
gauge bosons, containing at least eigth gluons.\par 
\smallskip
The data presented in 
Table 3  list the  properties of the four interactions. A new
entry here
is the value of the interaction radius. For the gravitational,
the weak, and the electromagnetic interaction, the magnitude of
the radius $r$ is determined from the uncertainty relation,
$\hbar /m_Bc,\; m_B$ denoting the mass of an exchange
particle. In the
case of the strong interaction, the interaction radius, $r_N$,
may
be regarded either as an empirical constant or as the distance
at
which the value of the coupling constant $\alpha_s$ becomes
unity.
\begin{center}
TABLE 3
\end{center}

\section{GRAVITATIONAL INTERACTION:
STRUCTURE OF THE MACROCOSMOS}

\subsection{Structural Elements of the Universe}

The structure of the macrocosmos is manifested on many
different scales, ranging from the universe on the largest
scale, down to galaxies, stars, and planets. Only objects of
the
microcosmos, such as quarks and leptons, may be devoid of
further substructure (However, a recent experiment is
suggesting that quarks and gluons may be composed of more
fundamental particles (Wilczek, 1996)). 
Most scales are
determined to an order of magnitude by a few physical
constants. In particular, the mass scale and length scale (in
units of the proton mass $m_p$ and the Bohr radius $a_0$) of
structures down to the planets can be expressed in terms of
the electromagnetic fine structure constant $\alpha_e$, eq.(4),
the
gravitational fine structure constant $\alpha_g$ eq.(3), and
the
electron-to-proton mass ratio ($m_p\approx 1837 m_e$).\par
\smallskip 
The following considerations are based on order of magnitude
arguments, factors of order unity (like $\pi$) being
neglected. These results have been published in the
astrophysical literature at several occasions and are collected
here for easy reference (a detailed discussion is contained in
Carr and Rees, 1979; Barrow and Tipler, 1988).\par 

\subsubsection{ The Planck scales} The only quantities of
dimensions mass
and length which can be constructed from $G, \hbar$ and $c$ are
the Planck scales:
\begin{equation}
M_{Pl}\sim(\frac{G}{\hbar c})^{-1/2} \sim
10^{-5}\;g,\;\;\;R_{Pl}\sim(\frac{G}{\hbar c^3})^{1/2} \sim
10^{-33}\;cm.
\end{equation}
Using the gravitational fine structure constant, 
eq.(3), these scales can be expressed as
\begin{equation}
M_{Pl} \sim \alpha_G^{-1/2} m_p,\;\;\;R_{Pl} \sim
\alpha_G^{1/2} r_p.
\end{equation}
Thus $M_{Pl}$ is much larger than $m_p$ but $R_{Pl}$ is much
smaller than $r_p$ ($r_p$ being the size of a proton that can
be taken to be the Compton wavelength associated with its rest
mass, $r_p\sim \hbar/m_pc\sim 10^{-13}\; cm;$ accordingly, the
corresponding timescale is $t_p \sim r_p/c \sim 10^{-23}\;s$).   
 The Planck length is the scale on which
quantum gravitational fluctuations in the metric become of the
order of unity, so the concept of space breaks down at such
small scales. $M_{Pl}$ can be interpreted as the mass of a
black hole of radius $R_{Pl}$. Space may be thought of as being
filled with virtual black holes of this size. Such
``instantons''
may play an important role in quantum gravity theory.

\subsubsection {The Universe} 
In the simplest Friedmann cosmological
model, the age of the universe $t_0$, is of the order of
$H^{-1}_0$ where $H_0$ is the Hubble parameter (this relation
fails only if the universe is closed and near its maximum
expansion). Since $H_0\sim 50\; km\;s^{-1}\;Mpc^{-1}$, this
implies
$t_0\sim 10^{10}$ yr, a conclusion which is supported by several
independent arguments. The associated horizon size (the
distance travelled by light since the birth of the
universe) satisfies the approximate relationship
\begin{equation}
ct_0 \sim \alpha_g^{-1}(\frac{\hbar}{m_ec})\sim
(\frac{\alpha_e}{\alpha_g})a_0,
\end{equation}
\noindent
where $a_0$ denotes the radius of the lowest energy electron
orbit of a hydrogen atom, $a_0\sim \hbar^2/m_ee^2\sim 10^{-8}\;
cm \sim 1\; Bohr.$  The ratio of the size of the observable
universe to the size of an atom is comparable with the ratio
of the electrical (or nuclear) and gravitational forces between
elementary particles. There is no explanation for this well
known coincidence within conventional physics, but Dirac 
(1937, 1938)  has
conjectured that $\alpha_g$ is always given by
\begin{equation}
\alpha_g\sim \frac{\hbar}{m_ec^2t}
\sim(\frac{t}{t_e})^{-1}.
\end{equation}
\noindent
Assuming that $\hbar, c,\; \mbox{and}\; m_e$ are constant in
time, this requires that $G$ decreases as $t^{-1}$, so Dirac
invokes eq.(12) as the basis for a new cosmology (Barrow, 1996).
Such a
variation of $G$ is inconsistent with current observations. The
total
mass associated with the observable universe (the
mass within the horizon volume) is $\sim\rho_0\;c^3\;t^3_0$,
where $\rho_0$ is the present matter density, given by the
Friedmann equation,
\begin{equation}
\rho_0 = \frac{3H^2_0}{8\pi G}+\frac{Kc^2}{16\pi G}.
\end{equation}
Here $K$ is the scalar curvature of the universe. Providing the
$K$ term is smaller than the others, one deduces that the mass
of
the universe is
\begin{equation}
M_u\sim c^3t^3_0G^{-1}H^2_0 \sim
\frac{c^3t_0}{G} \sim \alpha_g^{-2}(\frac{m_p}{m_e})m_p.
\end{equation}
\noindent
The fact that the number of protons in the universe is of the
order of $\alpha_g^{-2}$, is
thus explained providing one can justify neglecting the $K$
term in eq.(13). It has been argued that $K$ must always
be zero by appealing to Mach's principle but, apart from this,
there may be  reasons for expecting that the $K$ term
is small. If $K$ is negative, galaxies could not have condensed
out from the general expansion unless $(-K)$ were less than
$G\rho /c^2$ at their formation epoch. Otherwise, their
gravitational binding energy would not have been large enough
to halt their expansion. The term $(-K)$ may exceed $G\rho
/c^2$
at the present epoch, but not by a large factor. If $K$ is
positive, it must be
$<G\rho _0/c^2$, otherwise the whole universe would have
recollapsed before $t\simeq t_{MS}$, where $t_{MS}$ is the
lifetime of a typical main-sequence star, say, the Sun.
Relationships (11) and
(14) also mean that the universe has an optical depth of the
order of unity to electron scattering.
\subsubsection{Galaxies} 
It is not certain how galaxies form,
so any
estimate of their scale is very model dependent (Rees and
Ostriker, 1977; Silk, 1977; Sciama, 1953).  One can 
assume that galaxies originate from overdense regions in the
gaseous primordial material, and that they have a mass $M$ and
radius $R_B$ when they become bound. After binding,
motions in the protogalaxy randomize and equilibrate in the
gravitational field of the galaxy at a radius $\sim R_B/2$.
Thereafter, they will deflate on a cooling timescale, with a
virial temperature
\begin{equation}
T\sim \frac{GMm_p}{kR}.
\end{equation}
Providing $kT$ exceeds one Rydberg the dominant cooling
mechanism is bremsstrahlung and the associated cooling
timescale is
\begin{equation}
t_{cool}\simeq \frac{m_e^2c^3}{\alpha_e
e^4n}(\frac{kT}{m_ec^2})^{1/2}.
\end{equation}
The free-fall timescale is
\begin{equation}t_{ff}\sim(\frac{GM}{R^3})^{-1/2},
\end{equation}
and this exceeds $t_{cool}$ when $R$ falls below a
mass-independent value
\begin{equation}
R_g\sim\alpha_e^4\alpha_g^{-1}(\frac{m_p}{m_e})^{1/2}a_0,
\end{equation}
which, from a more precise calculation, is 75 kpc. Until a
massive cloud gets within this radius it will contract
quasi-statically and cannot fragment into stars. This argument
applies only if the mass is so high that
$kT_{virial}> \alpha_e^2m_ec^2$ at the `magic radius' $R_g$;
that
is,
\begin{equation}
M\geq M_g=\alpha_g^{-2}\alpha_e^5(\frac{m_p}{m_e})^{1/2}\simeq
10^{12}M_\odot.
\end{equation}
\noindent
Gas clouds of mass $<M_g$ cool more efficiently owing to
recombination, and can never be pressure supported. Thus, $M_g$
is a characteristic maximum galactic mass. Primordial clouds of
mass $<M_g$ are inhibited from fragmentation and may remain as
hot pressure-supported clouds. This type of argument can be
elaborated and made more realistic (White and Rees, 1978; Rees
and Ostriker, 1977; Silk, 1977; Sciama, 1953); but one still
obtains a
mass $\sim M_g$ above which any fluctuations are likely to
remain amorphous and gaseous, and which may thus relate to the
mass-scale of galaxies. The quantities $M_g$ and $R_g$ may thus
characterise
the mass and radius of a galaxy. These estimates is the least
certain. The properties of galaxies may
be a consequence of irregularities imprinted in the universe by
processes at early epochs.

\subsubsection{Stars} 
The virial theorem implies that the gravitational
binding energy of a star  must be of the order of its internal
energy. Its internal energy comprises the kinetic energy per
particle (radiation pressure being assumed negligible for the
moment) and the degeneracy energy per particle. The degeneracy
energy will be associated primarily with the Fermi-momentum of
the free electrons, $p\sim\hbar/d$, where $d$ is their average
separation. Provided the electrons are non-relativistic, the
degeneracy energy is $p^2/2m_e$, so the virial theorem implies
(Dyson, 1972; Weisskopf, 1975; Haubold and Mathai, 1994)
\begin{equation}
kT+\frac{\hbar^2}{2m_e d^2}\sim
\frac{GMm_p}{R}\sim(\frac{N}{N_0})^{2/3}\frac{\hbar c}{d}.
\end{equation}
Here $N$ is the number of protons in the star, $N_0\equiv
\alpha_g^{-3/2}$, and $R\sim  N^{1/3}d$ is its radius. As a
cloud collapses under gravity, eq. (20) implies that, for
large $d, T$ increases as $d^{-1}$. For small $d$, however, $T$
will reach a maximum
\begin{equation}
kT_{max}\sim(\frac{N}{N_0})^{4/3}m_ec^2,
\end{equation}
and then decrease, reaching zero when $d$ is
\begin{equation}
d_{min}\sim(\frac{N}{N_0})^{-2/3}r_e,
\end{equation}
where $r_e$ is the size of an electron, $r_e\sim\hbar/m_ec \sim
10^{-10}$ cm; accordingly the electron timescale is defined as
$t_e=r_e/c \sim 10^{-20}$ s. A collapsing cloud becomes a star
only if $T_{max}$ is high
enough for nuclear reactions to occur, that is $kT_{max}>qm_e
c^2$ where $q$ depends on the strong and electromagnetic
interaction constants and is $\sim 10^{-2}$. From eq.
(21), one therefore needs $N>0.1 N_0$. Once a star has ignited,
further collapse will be postponed until it has burnt all its
nuclear fuel. An upper limit to the mass of a star derives from
the requirement that it should not be radiation-pressure
dominated. Such a star would be unstable to pulsations which
would probably result in its disruption. Using the virial
theorem (that is, eq.(20) with the degeneracy term
assumed negligible) to relate a star's temperature $T$ to its
mass $M\sim Nm_p$ and radius $R$, the ratio of radiation
pressure to matter pressure can be shown to be
\begin{equation}
\frac{P_{rad}}{P_{mat}}\sim \frac{aT^4 R^3}{NkT}\sim
(\frac{N}{N_0})^2,
\end{equation}
so the upper limit to the  mass of a star is also $\sim N_0
m_p$. A more careful calculation shows that there is an extra
numerical factor of the order of 10, so one expects all
main-sequence stars to lie in the range $0.1<N/N_0<10$
observed. Only assemblages of $10^{56}-10^{58}$ particles
can turn into stable main sequence stars with hydrogen-burning
cores.
Less massive bodies held together by their own gravity can be
supported by electron `exclusion principle' forces at lower
temperatures (they would not get hot enough to undergo nuclear
fusion unless squeezed by an external pressure); heavier bodies
are fragile and unstable owing to radiation pressure
effects. The central temperature, $T$, adjusts itself so that
the
nuclear energy generation rate balances the luminosity, the
radiant energy content divided by the photon leakage time,
\begin{equation}L\simeq acT^4R^4/\kappa M,
\end{equation}
where $\kappa$ is the opacity. The appropriate $\kappa$
decreases as $M$ increases (electron scattering being the
dominant opacity for upper main-sequence stars) but the energy
generation increases so steeply with $T$ that $M/R$ depends only
weakly on $M$.

\subsubsection{White dwarfs and neutron stars} 

When a star has burnt all
its nuclear fuel, it will continue to collapse according to
eq.(20) and, providing it is not too large, it will end
up as a cold electron-degeneracy supported `white dwarf' with
the radius $R\propto M^{-1/3}$ indicated by eq.(22).
However, when $M$ gets so large that $kT_{max}\sim m_ec^2$, the
electrons will end up relativistic, with the degeneracy energy
being $pc$ rather than $p^2/2m_e$. Thus the degeneracy term in
eq.(20) acts as $d^{-1}$ instead of $d^{-2}$, and
consequently there is no $T=0$ equilibrium state. From
eq.(21) this happens if $M$ exceeds the mass
\begin{equation}
M_c\sim \alpha_g^{3/2}m_p\sim 1 M_\odot,
\end{equation}
which characterises stars in general. A more precise expression
for this critical value of $M$ (Chandrasekhar mass) is
$5.6\mu^2 M_\odot$, where $\mu$ is the number of free electrons
per nucleon (Chandrasekhar, 1935). For a star which has burned
all the way up to
iron, $\mu\approx 1/2$ and the limiting Chandrasekhar mass,
taking into account the onset of inverse $\beta$-decay when
electrons get relativistic, is $\sim 1.25 M_\odot$. Stars
bigger than $M_c$ will collapse beyond the white dwarf density
but may
manage to shed some of their mass in a supernova explosion. The
remnant core will comprise mainly neutrons, the electrons
having been squeezed onto the protons through the reaction
$p+e^-\rightarrow
n+\nu$, and this core, if small enough, may be supported by 
neutron-degeneracy pressure. The limiting mass for a neutron
star is more difficult to calculate than that for a white dwarf
because of strong interaction effects and because, from
eq.(21), with $m_e\rightarrow m_p$, the particles which
dominate the neutron star's mass are relativistic. The maximum
mass is still of the order of $1M_\odot$, however, and a
neutron star bigger than this must collapse to a black hole.
The maximum mass lies close to the intercept of a black hole
line, given by eq.(26), and the nuclear density line
$\rho \sim m_p/r^3_p$. The intricacies of the line which
bridges  white dwarf and neutron star regimes reflect the
effects of gradual neutronisation and strong
interactions. Stars on this bridge would be unstable and so are
not of immediate physical interest.\par
\smallskip
The above order-of magnitude arguments show why the effects of
radiation pressure and relativistic degeneracy both become
important for masses $>\alpha_g^{-3/2}m_p$. Note also that
general relativity is unimportant for white dwarfs because the
binding energy per unit mass is only $\sim(m_e/m_p)$ of $c^2$
at the Chandrasekhar limit.
\subsubsection{Black holes} The radius of a spherically
symmetrical
black hole of mass $M$ is
\begin{equation}R=\frac{2GM}{c^2}\sim
\alpha_g(\frac{M}{m_p})r_p.
\end{equation}
This is the radius of the event horizon, the region from within
which nothing can  escape, at least, classically. Black
holes larger than $1M_\odot$ may form from the collapse of
stars or dense star clusters. Smaller holes require much
greater compression for their formation than could arise in the
present epoch, but they might have been produced in the first
instants after the Big Bang when the required compression
could have occurred naturally. Such `primordial' black holes
could have any mass down to the Planck mass. In fact, Hawking
(1974) has shown that small black holes are not black at all;
because
of quantum effects they emit particles like a black body of
temperature given by 
\begin{equation}
k\theta\sim\frac{\hbar
c^3}{GM}\sim\alpha_g^{-1}(\frac{M}{m_p})^{-1}m_pc^2.
\end{equation}
This means that a hole of mass $M$ will evaporate completely in
a time
\begin{equation}
t_{evap}\sim\alpha^2_g(\frac{M}{m_p})^3 t_p(N(\theta))^{-1}.
\end{equation}
$N(\theta)$ is the number of species contributing to the
thermal radiation: For $k\theta<m_ec^2$ these include only
photons, neutrinos, and gravitons; but at higher temperatures
other species may contribute. The evaporation terminates in a
violent explosion. For a solar mass hole, this quantum radiance
is negligible: $\theta$ is only $10^{-7}$ K and $t_{evap}\sim
10^{64}$ yr. But for small holes it is very important. A Planck
mass hole has a temperature of $10^{32}$K and only survives for
a time $\sim R_{Pl}/c\sim 10^{-43}$ s. Those holes which are
terminating their evaporation in the present epoch are
particularly interesting. As the age of the universe is
$t_0\sim\alpha_g^{-1} t_e$, the mass of such holes would be
\begin{equation}
M_h\sim \alpha_g^{-2/3}(\frac{t_0}{r_p})^{1/3}
m_p\sim\alpha_g^{-1}m_p\sim10^{15} \;g, 
\end{equation}
and , from eq.(26), their radius would be $r_p$. The
corresponding temperature is $\sim10$ MeV: Low enough to
eliminate any uncertainty in $N(\theta)$ due to species of
exotic heavy particles.

\subsection{Evolution of the Universe}

The origin of the universe is governed by laws of physics which
are still unknown at the time of writing the Encyclopedia of
Applied Physics (see Fig.6).\par
\smallskip
\begin{center}
FIGURE 6.
\end{center}

At $t=10^{-43}\; sec, T=10^{32}\; K$: The strong, weak, and
electromagnetic forces may appear as unified into one
indistinguishable
force. This period is often referred to as the Grand
Unification epoch. During this epoch, there may have
been a very rapid, accelerating expansion of the universe
called ``inflation''. The inflation made the universe very
large and flat, but also produced ripples in the space-time it
was expanding.\par
\smallskip
At $t=10^{-34}\; sec, T=10^{27}\; K$: The strong force becomes
distinct
from the weak and electromagnetic forces. The universe is a
plasma of quarks, electrons, and other particles. Inflation
ends and the expanding universe coasts, gradually slowing its
expansion under the pull of gravity.\par
\smallskip
At $t=10^{-10}\; sec, T=10^{15}\; K$: The electromagnetic and
weak
forces
separate (see Fig.7). An excess of one part in a billion of
matter over
antimatter has developed. Quarks are able to merge to form
protons and neutrons. Particles have acquired substance.\par
\smallskip
\begin{center}
FIGURE 7.
\end{center}

At $t=1\; sec, T=10^{10}\; K$: Neutrinos decouple and  the
electrons and positrons annihilate, leaving residual
electrons but predominantly the cosmic background radiation as
the main active constituent of the universe.\par
\smallskip
At $t=3\; min, T=10^9\; K$: Protons and neutrons are able to bind
together to form nuclei since their binding energy is now
greater than the cosmic background radiation energy. A rapid
synthesis of light nuclei occurs - first deuterium (D), then
heavier elements, primarily helium ($^3He, ^4He)$ but up to
lithium nuclei $^7Li$ (Tytler, Fan, and Burles, 1996; 
Gloeckler and Geiss, 1996).
About 75 percent of the nuclei are hydrogen and 25 percent are
helium; only a tiny amount are other elements. The heavier
elements are later formed by nuclear burning stars.\par
\smallskip
At $t=3\times 10^5\; yr, T=3\times 10^3\; K$: Matter and the
cosmic
background
radiation decouple as electrons bind with nuclei to produce
neutral atoms. The universe becomes transparent to the cosmic
background radiation, making it possible for the Cosmic
Background Explorer satellite (COBE) to map this
epoch of last scattering (see Fig.8).\par
\smallskip
\begin{center}
FIGURE 8.
\end{center}

A $t=10^9\; yr, T=18\; K$: Clusters of matter have formed from
the
primordial ripples to form quasars, primordial stars, and
protogalaxies.
In the interior of stars, the burning of the primordial
hydrogen and helium nuclei synthesizes heavier nuclei such as
carbon, nitrogen, oxygen, and iron. These are dispersed by
stellar winds and supernova explosions, making new stars,
planets, and life possible.\par
\smallskip
At $t=15 \times 10^9\; yr, T=3\; K$: The present epoch is reached
(see Fig.9). 
Five
billion years earlier, the solar system condensed from the
remnants of earlier stars. Chemical processes have linked
atoms together to form molecules and then solids
and liquids. Man has emerged from the dust of stars to
contemplate the universe around him.\par
\smallskip
\begin{center}
FIGURE 9.
\end{center}

\subsection{Large-Scale Distribution of Matter}

The nearest large galaxy to the Milky Way is the `Andromeda
galaxy' which is about 670 kpc
away.  Its mass is $\sim 3\times 10^{11} M_\odot$ or perhaps as
much as $1.5\times 10^{12}M_{\odot}$ if it has a massive halo) 
and it has
a size of
$\sim 50\; kpc.$  Studies show that it is a spiral galaxy. The
Andromeda galaxy
 and the Milky Way lie nearly in each other's planes, but their
spins are opposite. It may be noted
that galaxies are packed in the universe in a manner very
different from the way the stars are distributed inside a
galaxy:   The distance from the Milky Way to the nearest large
galaxy is only 20 galactic diameters while the distance from
the Sun to the nearest star is thirty million times the
diameter of such individual stars. There is some evidence to
suggest that the Andromeda galaxy and the Milky Way are
gravitationally
bound to each other.  They have a relative velocity - towards
each other - of about $300\; km\; s^{-1}$.  These two are
only the two largest members of a group of more than 30
galaxies all of
which together constitute what is known as the `Local Group'.  
The entire Local Group is irregular in shape and  can be
contained within a spherical
volume of $\sim$ 2 Mpc in radius. This kind of clustering
of galaxies into groups is typical in the distribution of the
galaxies in the universe.  A study within a size of
about 20 Mpc from the Milky Way shows that only (10-20)\% of the
galaxies do not belong to any group; they are isolated
galaxies, called
`field' galaxies. Groups may typically consist of up to 100
galaxies; a system with more than 100 galaxies is
conventionally called a `cluster'. The sizes of groups range
from a few hundred kpc to 2 Mpc. 
Clusters have a size of typically a few Mpc. Just
like galaxies, one may approximate clusters and groups as
gravitationally bound systems of effectively point particles.
The large gravitational potential energy is counterbalanced by
the large
kinetic energy of random motion in the systems.  The line of
sight velocity dispersion in groups is typically $200\; km
\;s^{-1}$ 
while that in clusters can be nearly $1000\; km\; s^{-1}$.\par
There
are
several similarities between clusters of galaxies and stars in
an elliptical galaxy. For example, the radial distribution of
galaxies in a cluster can be adequately fitted by the $R^{1/4}$
law
with an effective radius $R_e\approx (1-2)h^{-1} Mpc\; (0.5 \leq
h \leq 1$ related to the Hubble constant).   About
10\% of all galaxies are members of large clusters. In
addition to galaxies, clusters also contain very hot
intracluster gas at temperatures $(10^7-10^8)$ K. The two large
clusters nearest to the Milky Way are the Virgo cluster and
the Coma cluster.  The Virgo cluster, located $\sim$ 17 Mpc
away, has a diameter of about 3 Mpc and contains more than 2500
 galaxies.  This is a prominent irregular cluster and
does not exhibit a central condensation or a discernible
shape. Virgo has an intracluster X-ray emitting gas with a
temperature of $\sim 10^8$ K, which has at least ten times the
visible mass of the cluster.  The Coma cluster, on
the other hand, is almost
spherically symmetric with a marked central condensation.  It
has an overall size of $\sim$ 3 Mpc while its central core is
$\sim$ 600 kpc in size. The core is populated with elliptical
and spheroidal galaxies with a density nearly thirty times larger
than the Local Group. These values are typical for large
clusters.  Coma is located at a distance of $\sim$ 80 Mpc and
contains more than 1000 galaxies. The
distribution of galaxies around the Local Group has been
studied extensively. It turns out that most of the galaxies
nearby lie predominantly in a plane - called the super-galactic
plane - which is approximately perpendicular to the
plane of the Milky Way galaxy. The dense set of galaxies in this
plane is called the Local Supercluster (radius $\sim$ 37 Mpc)  
 and the Virgo cluster
is nearly at the centre of this highly flattened disc-like
system.  The term `supercluster' is just used to denote
structures
bigger than clusters. Broadly speaking, the Local
Supercluster consists of three components:  About 20\%
of the brightest galaxies, forming the core, is the Virgo
cluster itself; another 40\% of galaxies lie in a flat
disc with two extended, disjoint groups of galaxies; the
remaining 40\% is confined to a small number of groups
scattered around.  Nearly 80\% of all matter in the
Local Supercluster lies in a plane . The Local Supercluster is
clearly expanding. Studies of distant
galaxies show that there are many superclusters in our
universe separated by large voids. The distribution of matter
seems to be reasonably uniform when observed at scales
bigger than about 100 $h^{-1}$ Mpc.  For comparison, note that
the size of the observed universe is about 3000 $h^{-1}$  Mpc. 
Thus, one may treat the matter distribution in the universe
to be homogeneous while dealing with the phenomena at scales
larger than 100 Mpc. The standard cosmological model is based
on this
assumption of ``large scale'' homogeneity (see Fig.10 and Table
4; Geller and Huchra, 1989).\par
\smallskip
\begin{center}
FIGURE 10.
\end{center}

\subsection{Morphology of Galaxies}

Galaxies range widely in their sizes, shapes, and masses; 
nevertheless, one may talk of a typical galaxy as something
made out of $\sim 10^{11}$ stars.  Taking the  mass
of a star to be that of the Sun, the luminous mass in a galaxy
is $\sim 10^{11}M_{\odot} \approx 2\times 10^{44}\;g$.   This
mass
is distributed in a
region with a size of $\sim$ 20 kpc.  Even though most galaxies
have a mass of $\sim (10^{10}-10^{12}) M_{\odot}$   and a size
of $\sim (10-30)$
kpc, there are several exceptions at both ends of this
spectrum.  
For example, `warf galaxies'  have masses in the range
$\sim (10^5-10^7) M_\odot$  and radii of $\sim (1-3)$ kpc. 
There are
also some `giant galaxies' with masses as high as $10^{13}
M_\odot$.
Galaxies exhibit a wide variety in their shapes as well and
are usually classified according to their morphology (Hubble's
classification of galaxies, for example). One may divide them
into `ellipticals' and `discs'.  
Ellipticals are smooth, featureless, distributions of stars,
ranging in mass from $10^8 M_{\odot}$  to $10^{13} M_{\odot}.$ 
The proportion of elliptical galaxies in a region depends
sensitively on the
environment.   They contribute $\sim 10\%$  of all
galaxies in low density regions of the universe but $\sim 40\%$
 in dense clusters of galaxies. The second major
type of galaxy is the `piral' (or  `disc') to which  the Milky
Way belongs.  Spirals have a prominent
disc, made of Population I stars and contain a significant
amount of gas and dust. The name originates from the distinct
spiralarms which exist in many of these galaxies. 
In low density regions of the universe, $\sim 80\%$
of the galaxies are spirals while only $\sim 10\%$ of
galaxies in dense clusters are spirals.  This is complementary
to the behavior of ellipticals. The stars in a disc galaxy are
supported against gravity by their systematic rotation
velocity. The rotational speed of stars, $v(R)$, at radius $R$
has the remarkable property that it
remains constant for large $R$ in almost all spirals; the
constant value is typically  200-300 $km\; s^{-1}$.  
Most discs also contain a spheroidal component of Population II
stars. The luminosity of the spheroidal component relative to
the disc correlates well with several properties of these
galaxies.  This fact has been used for classifying the disc
galaxies into finer divisions. The stars in a galaxy are not
distributed in completely uniform manner.  A typical galaxy
contains several smaller stellar systems, each containing
$\sim (10^2-10^6)$ stars.  These systems, usually called star
clusters, can be broadly divided into two types called `open
clusters' and `globular clusters'.  Open clusters consist of
$\sim (10^2-10^3)$ Population I stars bound within a radius of
$\sim (1-10)$pc.
 Most of the stars in these clusters are quite
young.  In contrast, globular clusters are Population II
systems with $\sim (10^4-10^6)$ stars. The Milky Way contains 
$\sim 200$
globular clusters which are distributed in a spherically
symmetric manner about the centre of the galaxy.  Unlike open
clusters, the stars in globular clusters are quite old.  The
number density of stars in the core of a globular cluster,
$\sim 10^4 M_\odot\; pc^{-3}$, is much higher than that of a
typical
galaxy, $\sim (0.05 M_\odot\; pc^{-3}$.   The core radius of the
globular
clusters is $\sim$ 1.5 pc while the `tidal radius', which is
the radius at which the density drops nearly to zero, is $\sim$
50 pc. In addition to the stars, the Milky Way also contains
gas
and dust which contribute $\sim (5-10) \%$ of its mass. 
This interstellar medium may be roughly divided into a very
dense, cold, molecular component (with about $10^4$ particles
per
$cm^{-3}$ and a temperature of $\sim$ 100 K), made of
interstellar clouds, a second component which is atomic
but neutral, (with $n \approx 1\; cm^{-3}$ and $T \approx 10^3\;
K)$
and a third
component which is ionized and very hot (with $n\approx 10^{-3}
\;cm^{-3}$
and $T\approx 10^6\; K$).  Though the interstellar medium is
principally
made of hydrogen, it also contains numerous other chemical
species  and an
appreciable quantity of tiny solid particles (`dust'). The
spiral arms are concentrations of stars and interstellar gas
and are characterized by the presence of ionized hydrogen.  
This is also the region in which young stars are being formed.
For a more detailed study of galaxies, one can divide
ellipticals and spirals into subsets and also add two more
classes of galaxies, called `lenticulars' and `irregulars'. 
The ellipticals are subdivided as (E1, . . ., En, . . .) where
$n = 10(a-b)/a$ with $a$ and $b$ denoting the major and minor
axis
of the ellipticals. The `lenticulars' (also called SO) are the
galaxies `in between' ellipticals and spirals. They have a
prominent disc which contains no gas, dust, bright young stars
or spiral arms. Though they are smooth and featureless like
ellipticals, their surface brightness follows the exponential
law of the spirals. They are rare in low density regions (less
than 10\%) but constitute nearly half of the galaxies
in the high density regions. The spirals are subdivided into
Sa, Sb, Sc, Sd with the relative luminosity of the spheroidal
component decreasing along the sequence. The amount of gas
increases and the spiral arm becomes more loosely spiraled
along the sequence from Sa to Sd.  The Milky Way is between the
types Sb and
Sc. There also exists another class of spirals called `barred
spirals' which exhibit a bar-like structure in the centre.
They are classified as SBa, SBb etc. Finally, irregulars are
the galaxies which do not fall in the above mentioned
morphological classification. These are low luminosity, gas
rich systems with massive young stars and large HII regions
(i.e. regions containing ionized hydrogen). More than one
third of the galaxies in our neighborhood are irregulars. They
are intrinsically more difficult to detect at larger distances
because of their low luminosity. We have seen earlier that as
the stars evolve, their luminosity and hence the color
changes. Since galaxies are made of stars, galaxies will also
exhibit color evolution. Besides, the gas content and
elemental abundances of the galaxies will change as the stars
are formed and end as planetary nebulae or supernovae. 

\subsection{Quasars}                     

Most galaxies which are observed have a fairly low redshift $(z
<
0-5)$ and an extended appearance on a photographic plate. There
exists another important class of objects, called `quasars',
which exhibit large redshifts (up to $z\approx 5$) and appear
as
point sources in the photographic plate (Shaver et al., 1996).
Estimating the
distance from the redshift and using the observed luminosity,
it is found that quasars must have a luminosity of about
$L_q\approx
(10^{46}-10^{47})\; erg\; s^{-1}$. It is possible to estimate the
size of
the region emitting the radiation from the timescale in which
the radiation pattern is changing.  It
turns out that the energy from the quasar is emitted from a
very compact region. One can easily show that nuclear fusion
cannot be a viable energy source for quasars. It is generally
believed that quasars are fuelled by the accretion of matter
into a supermassive black hole $(M\approx 10^8 M_\odot)$ in the
centre of the host galaxy. The friction in the accretion disc
causes the
matter to lose angular momentum and hence spiral into the
black hole; the friction also heats up the disc. Several
physical processes transform this heat energy into radiation
of different wavelengths. Part of this energy can also come
out in the form of long, powerful, jets. The innermost regions
of the quasar emit X- and $\gamma$-rays.  Outer shells emit UV,
optical, and radio continuum radiation in the order of
increasing radius. Very bright quasars have an apparent
magnitude of $m_B\approx 14$ (which corresponds to a flux of
$10^{-25}\;
erg\;s^{-1}\; cm^{-2}\; Hz^{-1})$ while the faintest ones have
$m_B
\approx 23$. The
absolute magnitudes of the quasars are typically $-30 < M_B <
-23$; in contrast, galaxies fall in the band $-23 < M_B < -16$.
The relation between $m$ and $M$ for any given quasar depends on
the estimated distance to the quasar, which in turn depends on
the cosmological model and the quasar's redshift. The
luminosity function of the quasars, as a function of their
redshift, has been a subject of extensive study. These
investigations show that: (i) the space density of bright
galaxies (about $10^{-2}\; Mpc^{-3}$) is much higher than that of
quasars (about $10^{-5}\; Mpc^{-3}$) with $z < 2$ and (ii) bright
quasars
were more common in the past $(z\approx 2)$ than today
$(z\approx 0)$.
Quasars serve as an important probe of the high redshift
universe. Quasars are believed to
be one extreme example of a wide class of objects called
`active galaxies'. This term denotes a galaxy
which seems to have a very energetic central source of energy. 
This source is most likely to be a black hole powered by
accretion. One kind of active galaxy which has been studied
extensively are radio galaxies. The most interesting feature
about these radio galaxies is that the radio emission does not
arise from the galaxy itself but from two jets of matter
extending from the galaxy in opposite directions. It is
generally believed that this emission is caused by the
synchrotron radiation of relativistic electrons moving in the
jets. The moving electrons generate two elongated clouds
containing magnetic fields which, in turn, trap the electrons
and lead to the synchrotron radiation.

\subsection{Stellar Evolution}

The time evolution of a star, which can
be depicted as a path in the Hertzsprung-Russell diagram, is
quite complicated
because of many physical processes which need to be
taken into account. Detailed calculations, based on the
numerical integration of the relevant differential equations,
have provided
a fairly comprehensive picture of stellar evolution.  
One of the
primary sources of stellar energy is a series of nuclear
reactions converting four protons into a helium nucleus. 
Since the simultaneous collision of four particles is
extremely improbable, this process of converting hydrogen into
helium proceeds through two different sequences of intermediate
reactions, one called proton-proton chain and the other called
carbon-nitrogen-oxygen  cycle. In the p-p chain, helium is formed
through
deuterium and $^3He$ in the intermediate steps; this reaction
is
the dominant mechanism for hydrogen-helium conversion at
temperatures
below $\sim 2\times 10^7$ K.  In the CNO cycle, hydrogen
is
converted into helium through a sequence of steps involving
$^{12}C$ as a catalyst, i.e., the amount of $^{12}C$ remains
constant at
the end of the cycle of reactions. Since the Coulomb barrier
for carbon nuclei is quite high, the CNO cycle is dominant
only at higher temperature. The evolution of a star  like the
Sun  during the phase of hydrogen-helium conversion, called the
`main
sequence' phase, is fairly stable. The
stability is essentially due to the following regulatory
mechanism: Suppose the temperature decreases slightly causing
the nuclear reaction rate to decrease. This will make gravity
slightly more dominant, causing a contraction. Once the star
contracts, the temperature will again increase, thereby
increasing the rate of nuclear reactions and the pressure
support. This will restore the balance. After the burning of
core hydrogen ends, the core will undergo a contraction,
increasing its temperature; if the star now heats up beyond
the helium ignition temperature, then the burning of helium
will start and stabilize the star. In principle, such a
process can continue with the building up of heavier and
heavier elements. But to synthesize elements heavier than He
is not straight forward because He has a very high binding energy
per
nucleon among the light elements. Stars achieve synthesis of
post-helium elements through a process known as `triple-alpha
reaction' which proceeds as $^4He(^4He, ^8Be)\gamma;\; 
^8Be(^4He,^{12}C)\gamma$. 
Once $^{12}$C has been synthesized, production of heavier
elements 
like $^{16}O,\; ^{20}Ne,\; ^{24}Mg$ etc. can
occur through various channels, provided temperatures
are high enough, and the star can evolve through successive
stages of nuclear burning. The ashes of one stage can become
the fuel for the next stage as long as each ignition
temperature is reached. Such stars will evolve into
structures which contain concentric shells of elements. For
example, a $15 M_{\odot}$  star, during its final phase can have
layers
of iron, silicon, oxygen, neon, carbon, helium, and hydrogen
all burning at their inner edges. \par
The details of the above
process, which occurs after the exhaustion of most of the fuel
in the core, depend sensitively on the core mass. Consider, for
example, a star with $M > 1M_\odot$ .  Its evolution proceeds
in the
following manner:  Once the hydrogen is exhausted in the core,
the core, containing predominantly helium, undergoes
gravitational
contraction. This increases the temperature of the material
just beyond the core and causes renewed burning of hydrogen in
a shell-like region. Soon, the core contracts rapidly,
increasing the energy production and the pressure in the
shell, thereby causing the outer envelope to expand. Such an
expansion leads to the cooling of the surface of the star. 
About this time, convection becomes the dominant mechanism for
energy transport in the envelope and the luminosity of the
star increases due to  convective mixing. This is usually
called the `red-giant' phase. During the core contraction, the
matter gets compressed to very high densities ($\sim 10^5\; g\;
cm^{-3}$) so that it behaves like a degenerate gas and not as
an
perfect gas.  Once the core temperature is high enough to
initiate the triple-alpha reaction, helium burning occurs at
the core.  Since degenerate gas has a high thermal
conductivity, this process occurs very rapidly, called `helium
flash'.  If the core was dominated by gas pressure, such an
explosive ignition would have increased the pressure and led
to an expansion; but since the degeneracy pressure is
reasonably independent of temperature, this does not happen. 
Instead the evolution proceeds as a run-away process: The
increase in temperature causes an increase in the triple-alpha
reaction rate, causing further increase in temperature etc.
Finally, when the temperature becomes  $\sim 3.5 \times 10^8$
K,
the
electrons become non-degenerate; the core then expands and
cools. The star has now reached a stage with helium burning in
the
core and hydrogen burning in a shell around the core. Soon the
core
is mostly converted into carbon and the reaction again stops. 
The process described above occurs once again, this time with
a carbon-rich, degenerate core, and a helium burning shell. This
situation, however, turns out to be unstable because the
triple-alpha reaction is highly sensitive to temperature. This
reaction can over respond to any fluctuations in pressure or
temperature thereby causing pulsations of the star with
increasing amplitude. Such violent
pulsation can eject the cool, outer layers of the star leaving
behind a hot core. The ejected envelope becomes what is known
as a `planetary nebula'. The above discussion assumes that the
core could contract sufficiently to reach the ignition
temperature for carbon burning. In low mass stars, degeneracy
pressure stops the star from reaching this phase and it ends
up as a `white dwarf' supported by the degeneracy pressure of
electrons against gravity. A more complicated sequence of
nuclear burning is possible in stars with $M>> 1M_\odot$. After
the
exhaustion of carbon burning in the core, one can have
successive phases with neon, oxygen and silicon burning in the
core with successive shells of lighter elements around it.
This process can proceed until $^{56}Fe$ is produced in the core.

The binding energy per nucleon is maximum for the $^{56}Fe$
nucleus; hence it will not be energetically feasible for
heavier elements to be synthesized by nuclear fusion. The core
now collapses violently reaching very high, about
$10^{10}
K$, temperatures.  The $^{56}Fe$ photo-disintegrates into alpha
particles, and then even the alpha particles disintegrate at such
high temperatures to become protons. The collapse of the core
squeezes
together protons and electrons to form neutrons and the
material reaches near-nuclear densities forming a `neutron
star'. There exist several physical processes which can
transfer the gravitational energy from core collapse to the
envelope, thereby leading to the forceful ejection of the
outer envelope causing a `supernova explosion'. A remnant
with smaller mass is left behind. Numerical studies show that
stars with $M > 8M_{\odot}$  burn hydrogen, helium, and carbon
and evolve
rather smoothly. During the final phase, such a star explodes
as a supernova leaving behind a remnant which could be a white
dwarf, neutron star, or black hole. It is generally believed
that stars with masses in the intermediate range, $2M_{\odot}
< M <
8M_{\odot}$, do not burn hydrogen and helium in degenerate
cores but
evolve through carbon burning in degenerate matter (for $M >
4M_{\odot}$)
ending again in a supernova explosion.  Stars with lower
mass do not explode but end up as planetary nebulae. Low mass
stars with $M < 2 M_\odot$  ignite helium in degenerate
cores at the
tip of the red giant branch and then evolve in a complicated
manner. The synthesis of elements as described above proceeds
smoothly up to $^{56}Fe$. Heavier elements are formed by nuclei
absorbing free neutrons, produced in earlier reactions by
two different processes called the `r-process' (rapid neutron
capture process)
and the `s-process' (slow neutron capture process). During the
supernova
explosion, a significant part of the heavy elements
synthesized in the star will be ejected out into the
interstellar space. A second generation of stars can form from
these gaseous remnants. The initial composition of material in
this second generation will contain a higher proportion of
heavier elements (called `metals') compared to
the first generation stars.  Both these types of stars are
observed in the universe; because of historical reasons, stars
in the second generation are called population I stars while
those in the first generation are called population II stars.\par
The above discussion shows how stars could synthesize heavier
elements, even if they originally start as gaseous spheres
of hydrogen.  The study of the spectra of stars allows  to
determine the relative proportion of various elements present
in the stars. Such studies show that population II stars are
made of about 75\% hydrogen and 25\% helium; 
even population I stars consist of an almost similar
proportions of hydrogen and helium with a small percentage of
heavier elements ($\sim 2\%$). It is possible to show, using
stellar
evolution calculations, that it is difficult for the stars to
have synthesized elements in such a proportion, if they
originally had only hydrogen (called primordial or population III
stars). Hence, such a universal
composition leads us to conjecture that the primordial gas
from which population II stars have formed must have been
a mixture of hydrogen and helium in the ratio 3:1 by weight. 
Heavier elements synthesized by these population II stars
would have been dispersed in the interstellar medium by
supernova explosions. The population I stars are supposed to
have been formed from this medium, containing a trace of
heavier elements. The helium present in the primordial gas
should have been synthesized at a still earlier epoch, the Big
Bang, and
cannot be accounted for by stellar evolution. Cosmological
models provide an
explanation for the presence of this primordial helium.\par
\smallskip
\begin{center}
TABLE 4.
\end{center}

\section{MATHEMATICS AND PHYSICS AND THE\\
STRUCTURE OF THE UNIVERSE}

\subsection{General Relativity and Quantum Field Theory}

A mathematician represents the motion of the planets of the
solar system by a flow line of an incompressible fluid in a
54-dimensional phase space, whose volume is given by the
Liouville measure; while a famous physicist is said to have
made the statement that the whole purpose of physics is to
find a number, with decimal points, etc.!, otherwise you have
not done anything (Manin, 1980). Despite such extreme points
of view it can be safely said that the relations between
mathematics and physics (and astronomy) have been very
productive in past epochs of the evolution of these
disciplines (Barrow and Tipler, 1988).\par 
\smallskip
Mathematical structures entered the development of physics,
and problems emanating from physics influenced developments in
mathematics. Examples are the role of differential geometry in
general relativity, the dynamical theory of space and time,
and the influence of quantum mechanics in the development of
functional analysis and built on the understanding of Hilbert
spaces. A prospective similar development occurred only
recently when non-Abelian gauge theories emerged as the
quantum field theories for describing fundamental particle
interactions. Yang-Mills theory found its mathematical
formulation in the theory of principal fiber bundles. The
understanding of anomalies in gauge theories involved the
theory of families of elliptic operators and representation
theory of infinite-dimensional Lie algebras and their
cohomologies. Based on these developments, there are two
fundamental theories in modern physics: General relativity and
quantum field theory. General relativity describes
gravitational forces on an astronomical scale. Quantum field
theory describes the interaction of fundamental particles,
electromagnetism, weak, and strong forces. The formal
quantization of general relativity is leading to infinities
and the current understanding is that the mathematical
machinery is missing to accomplish the unification of general
relativity and quantum field theory. More recently attention
has turned to the exploration of the mathematical structure of
non-Abelian gauge theories and to the more ambitious attempts
to construct unified theories of all the fundamental
interactions of matter together with gravity. A success of
such attempts may reveal new insights for structural elements
of the microcosmos and macrocosmos (Schmid, 1992).

\subsection{Macroscopic Dimension of Space}

Of all the fundamental constants the most familiar one is the
dimension of physical space, $N=3$. Variation of such a
fundamental characteristic as the dimension N may lead to
unpredictable changes of physical laws. It was Paul Ehrenfest
(1880-1933) who,
in 1917, was trying to answer the question of why physical
space is three-dimensional. From physics, we are familiar with
the analogy between Coulomb's law and Newton's law (eqs.(1) and
(2)). In both
cases, the force is $F\propto r^{-2}$. In physics these laws
are treated
separately. This lack of
coherency obscures a profound relationship of the
electromagnetic and the gravitational forces with the
properties of space, in particular, with its dimension. Two
properties are common to the gravitational and the
electromagnetic interactions: Both are weak and long-range. In
modern language, this means that the mass of the gauge boson is
zero in both cases, implying that the interaction
radius is infinite and that the interaction constants are
small, $\alpha_g$, $\alpha_e<<1$ (eqs. (3) and (4)). In the
language of physics
these properties mean, that the lines of forces, originating at
the point where their source is
located, run to infinity, not intersecting with each other,
provided that no other source is present. The fact that the
lines of force extend to infinity reflects the long-range
character of the gravitational and electromagnetic forces; the
absence of intercepts signifies that there is no reciprocal
action between the lines of force, i.e., that the interactions
under consideration are weak. The combination of both
properties, the weakness and the long-range action, is not
characteristic for the other interactions. The force $F$ exerted
by one
particle on another particle, a distance $r$ apart, is
proportional to the density $n_i$ of the lines of force.
Accordingly, 
\begin{equation}
F\propto n_i=\frac{f}{4\pi r^2}.
\end{equation}
The proportionality constant $f$ in (30) is by definition equal
to the product of the charges of both
particles (Coulomb's law) or of their masses (Newton's law).
The denominator in (30) gives the surface area $S$ of a sphere of
radius $r$. For
the three-dimensional space, this quantity equals $S_3 = 4\pi 
r^2=a_3r^{3-1}$. These considerations can be repeated for the
more general
case of an N-dimensional space. The surface area of a sphere
in such a space is $S_N =a_Nr^{N-1}$. Therefore, the force
$F_N$, acting
in an N-dimensional space, is 
\begin{equation}
F_N=\frac{b_N}{r^{N-1}}.
\end{equation}
Accordingly, the potential energy has the form 
\begin{equation}
U_N=\frac{-b_N}{[(N-2)r^{N-2}]},
\end{equation}
\noindent
where $N\not=2$; for $N=2$, the dependence is logarithmic. It
should
be stressed that these expressions apply only for integer $N$,
and only for long-range forces in the quasistatic approximation,
i.e., for motion in a central force field. The existence of
stable orbits in a central force field in an $N$-dimensional
space is determined by (31) and consequently by the dimension
N. From mechanics, it is
known that the existence of stable orbits depends on the form
of the $r$-dependence of the effective potential
\begin{equation}
U_{Ne}=U_N+\frac{M^2}{2mr^2},
\end{equation}
\noindent
where $M^2/(2mr^2)$ is the centrifugal energy, $M$ the angular
momentum, and $m$ the mass of the body moving at a distance $r$.
A
stable state is possible if the dependence $U_{Ne}(r)$ 
has a minimum at a value of $r$ different from zero or infinity.
Henceforth the attractive forces are considered for which 
$U_N<0$. The results of an analysis of the function $U_{Ne}(r)$
with
regard to the existence of an extremum, are the following:\par
\medskip
For $N>4$, the dependence $U_{Ne}(r)$ has a maximum at $r\not=
0$ and a
minimum at $r=0$ which corresponds to merging of the two
particles.\par
For $N=4$, the dependence $U_{Ne}(r)$ is given by a
monotonically decreasing function exhibiting no extrema.\par
     For $N=2$ and $N=3$, the function $U_{Ne}(r)$ has a
minimum at $r\not=0$
and $r \not= \infty$.\par
     For $N=1$, the function $U_{Ne}(r)$ is monotonically
increasing.\par
\medskip
The existence of a minimum in the dependence $U_{Ne}(r)$ is a
necessary condition for the stability of motion. For this
reason, the existence and the properties of closed orbits are
determined by the dimension of space, $N$. For $N>4$, there is
no
minimum at $r\not=0,\infty$ ; consequently there are no stable
closed
orbits. Any motion, caused by long-range forces would be of
one of the following two types: Either it is infinite (the
body escapes to infinity) or otherwise, the moving body falls
on a massive central body. At $N=2$ or $N=3$, all types of
motion
are possible: Infinite motion, fall onto a central body and,
notably, motion in stable, closed orbits. For $N=1$, only
finite
motion is possible; a body cannot escape to infinity. In the
one-dimensional case, no orbital motion occurs, and the
centrifugal potential is zero $(M=0)$. The effective potential,
(32), is then $U_1=b_1r$, and the force $F_1$=const. This
corresponds
to an
infinitely deep potential well. To remove the body from this
well, an infinitely large force has to be applied; this means
it would be impossible for the body to escape to infinity.
Hence, the degree of stability grows as the dimension N
decreases. For $N>4$, there are no analogues to planetary
systems. Similar considerations in the framework of quantum
mechanics have demonstrated that for $N>4$, stable atomic
systems do not exist, either. It appears that the absence of
analogues of planets and atoms for $N>4$ is a clue to the
understanding of the significance of the space dimension
$N=3$, with regard to the existence of structural elements in the
universe.

\subsection{Microscopic Dimension of Space}

In classical physics, particles have definite locations and
follow exact trajectories in spacetime. In quantum mechanics,
wavepackets propagate through spacetime, their positions and
velocities uncertain according to Heisenberg's uncertainty
principle. In string theory, point particles are replaced by
tiny loops with the result that the concept of spacetime
becomes ``fuzzy'' at scales comparable to the square root of a
new fundamental constant, $\alpha'\approx (10^{-32} cm)^2$,
introduced as
string tension in string theory. Employing both string tension
($\alpha'\not= 0$) and  quantum effects ($\hbar \not= 0$) leads
to results that may
change the conventional notion of spacetime (Donaldson, 1996).
The situation of introducing $\alpha`$ in string theory is
similar to passing from classical to quantum mechanics by the
introduction of Planck`s constant $\hbar$. In string theory the
one-dimensional trajectory of a particle in spacetime is replaced
by a two-dimensional orbit of a string. According to Heisenberg`s
uncertainty principle, at a momentum $p$ one can probe a distance
$x\approx \hbar /p$. However, the introduction of $\alpha`s $
acts, as if Heisenberg`s uncertainty principle would have two
terms, $\Delta x\geq \hbar /\Delta p + \alpha` (\Delta p/\hbar)$,
where the second term reflects the fuzziness due to string
theory. Then, the constant $\alpha`$ would be the absolute
minimum uncertainty in length in any physical experiment. The
consequences of this type of quantum field theory for fundamental
particles and the structure of the universe are still to be
discovered. 
\clearpage
\noindent
GLOSSARY (of terms related to the principal structural elements
of the universe)\par
\smallskip
\noindent
Antiparticle: A particle of opposite charge but otherwise
identical to its
partner. Most of the observable universe consists of particles
and matter,
as opposed to antiparticles and antimatter.\par
\smallskip
\noindent
Asteroids: Small planetlike bodies of the solar system.\par
\smallskip
\noindent
Atom: The basic building block of matter. Each atom consists of a
nucleus
with positive electric charge and a surrounding cloud of
electrons with
negative charge.\par
\smallskip
\noindent
Baryon: Type of hadron, consisting of proton, neutron, and the
unstable
hyperons (and their antiparticles).\par
\smallskip
\noindent
Binary System: A system of two objects orbiting around a common
center. The
objects may be stars or black holes or a star and a black
hole.\par
\smallskip
\noindent
Black Hole: A region in which matter has collapsed to such an
extend that
light can no longer escape from it.\par
\smallskip
\noindent
Boson: A class of particles with integer units of the basic unit
of spin
$h/2\pi$.\par
\smallskip
\noindent
Cluster of Galaxies: An aggregate of galaxies. Clusters may range
in
richness from loose groups, such as the Local Group, with 10 to
100
members, to great clusters of over 1000 galaxies.\par
\smallskip
\noindent
Cluster of Stars (Globular Star Cluster): An aggregate of stars.
Globular
star clusters contain the oldest stars in the galaxy and
high-velocity
stars.\par
\smallskip
\noindent
Comet: A diffuse body of gas and solid particles that orbit the
Sun in a
highly eccentric trajectory.\par
\smallskip
\noindent
Cosmic Microwave Background Radiation: Diffuse isotrope radiation
whose
spectrum is that of a blackbody at 3 degrees kelvin and
consequently is
most intense in the microwave region of the spectrum.\par
\smallskip
\noindent
Cosmic String: A hypothetical one-dimensional, string-like object
that is
made from a curvature of space.\par
\smallskip
\noindent
Cosmogony: The study of the origin of celestial systems, ranging
from the
solar system to stars, galaxies, and clusters of galaxies.\par
\smallskip
\noindent
Cosmology: The study of the large-scale structure and evolution
of the
universe.\par
\smallskip
\noindent 
Dark Matter: Matter whose presence is inferred from dynamical
measurements
but which has no optical counterpart.\par
\smallskip
\noindent
Electron: A particle of matter with negative electric charge. All
chemical
properties of atoms and molecules are determined by the
electrical
interactions of electrons with each other and with the atomic
nuclei.\par
\smallskip
\noindent
Fermion: A particle with half integral units of the basic unit of
spin,
$h/2\pi$.\par
\smallskip
\noindent
Galactic Nucleus: Innermost region of a galaxy exhibiting a
concentration
of stars and gas.\par
\smallskip
\noindent
Galaxy: A large gravitationally bound cluster of stars that all
orbit
around a common center. Galaxies are basic building blocks of the
universe.\par
\smallskip
\noindent
Graviton: The particle which, according to wave-particle duality,
is
associated with gravitational waves.\par
\smallskip
\noindent
Group of Galaxies: Gravitationally bound system of few
galaxies.\par
\smallskip
\noindent
Group of Stars: Gravitationally bound system of few stars
(multiple stars).\par
\smallskip
\noindent
HI Cloud: Cloud of cool, neutral hydrogen.
\par
\smallskip
\noindent
HII Region: Cloud of hot, ionized hydrogen, usually heated by a
massive
young hot star.\par
\smallskip
\noindent
Hadron: Particle (protons, neutrons, mesons) which takes part in
strong
nuclear reactions.\par
\smallskip
\noindent
Halo: The diffuse, nearly spherical cloud of old stars and
globular
clusters that surrounds a spiral galaxy.\par
\smallskip
\noindent
Hertzsprung-Russell Diagram: Plot of stellar luminosity (or
absolute
magnitude) against effective temperature (or color), in which the
evolution
of stars of different masses may be followed.\par
\smallskip
\noindent
Hubble Classification of Galaxies: Organizes galaxies according
to shape.
They range from amorphous, relatively uniform elliptical systems
to highly
flattened spiral disks with prominent nuclei. It is not a
classification
based on evolution but one of different rates of star
formation.\par
\smallskip
\noindent
Hypergalaxy: A system consisting of a dominant spiral galaxy
surrounded by
a cloud of dwarf satellite galaxies, often ellipticals. The Milky
Way and
the Andromeda galaxy are hypergalaxies.\par
\smallskip
\noindent
Irregular Galaxy: A galaxy without spiral structure or smooth,
spheroidal
shape, often filamentary or very clumpy, and generally of low
mass ($10^7$
to $10^{10} M\odot$).\par
\smallskip
\noindent
Intergalactic Gas: Matter that is present in the region between
galaxies.
It has been detected in considerable amounts in great clusters of
galaxies,
where the intergalactic gas is so hot that it emits copious
amounts of x-
rays.\par
\smallskip
\noindent
Interstellar Grains: Small needle-shaped particles in the
interstellar gas
with dimensions from $10^{-6}$ to $10^{-5}$ cm. They are
primarily composed
of silicates and strongly absorb, scatter, and polarize visible
light in
the far-infrared region of the spectrum.\par
\smallskip
\noindent
Lepton: Particles (neutrinos, electrons, muons) which do not take
part in
strong interactions.\par
\smallskip
\noindent
Local Group: Small group of 30 or so galaxies of which the Milky
Way and
the Andromeda galaxy are the two dominant members.\par
\smallskip
\noindent
Meson: A class of strongly interacting particles with zero baryon
number,
among them the pi meson.
\par
\smallskip
\noindent
Meteorites: A solid portion of a meteoroid that has reached the
Earth's
surface.\par
\smallskip
\noindent
Molecular Cloud: An interstellar cloud consisting predominantly
of
molecular hydrogen, with trace amounts of other molecules such as
carbon
monoxide and ammonia.\par
\smallskip
\noindent
Molecule: An entity made of several atoms that share their
electron clouds
with each other. \par
\smallskip
\noindent
Neutrino: A particle that resembles the photon, except that it
interacts
weakly with matter. Neutrinos come in at least three varieties,
known as
electron-type, muon-type, and tauon-type.\par
\smallskip
\noindent
Neutron: The uncharged particle found along with protons in
atomic nuclei.\par
\smallskip
\noindent
Neutron Star: Cold, degenerate, compact star in which nuclear
fuels have
been exhausted and pressure support against gravity is provided
be the
pressure of neutrons.\par
\smallskip
\noindent
Nucleon: Nuclear particles, i.e. neutrons and protons.\par
\smallskip
\noindent
Observable Universe: The extend of the universe that we can see
with the
aid of large telescopes. Its ultimate boundary is determined by
the horizon
size.\par
\smallskip
\noindent
Photon: A discrete unit of electromagnetic energy. The particle
which,
according to wave-particle duality, is associated with
electromagnetic
energy.\par
\smallskip
\noindent
Positron: The antiparticle of an electron, having positive charge
but being
otherwise similar.\par
\smallskip
\noindent
Proton: The positively charged particle found along with neutrons
in atomic
nuclei\par
\smallskip
\noindent

Pulsar: A magnetized, spinning neutron star that emits a beam of
radiation;
radio waves and sometimes also light and X-rays).\par
\smallskip
\noindent
Quark: Particles of which all hadrons are supposed to be
composed.\par
\smallskip
\noindent
Quasar: Luminous and compact quasi-stellar radio source related
to violent
events in the nuclei of a galaxy, believed to be powered by a
massive black
hole.\par
\smallskip
\noindent
Red Giant: Phase in star's evolution after completion of hydrogen
burning
when outer layers become very extended.\par
\smallskip
\noindent
Spiral Galaxy: A galaxy with a prominent nuclear bulge and
luminous spiral
arms of gas, dust, and young stars that wind out from the
nucleus.\par
\smallskip
\noindent 
Star: Stars are basic building blocks of the universe.\par
\smallskip
\noindent 
Supercluster: A cluster of clusters of galaxies.\par
\smallskip
\noindent 
Universe: Our universe.\par
\smallskip
\noindent 
White Dwarf: Cool, degenerate, compact star, in which nuclear
fuels are
exhausted and pressure support against gravity is provided by the
degeneracy pressure of electrons.
\clearpage
\noindent
{\large\bf List of Works Cited}\par
\bigskip
\noindent
Alpher, R.A., Herman, R.C. (1949),  Physical Review 75,
1089-1095.\par
\noindent
Alpher, R.A., Herman, R.C. (1950),  Reviews 
of Modern Physics 22, 153-219.\par
\noindent
Alpher, R.A., Follin Jr., J.W., Herman, R.C. (1953), Physical
Review 92, 1347-1361.\par
\noindent
Barnett, R.M. et al. (1996), Review of Particle Physics: Particle
Data Group,\par 
Physical Review, D54, 1-720.\par
\noindent
Barrow, J.D. (1996), Monthly Notices of the Royal Astronomical
Society, 282,\par 
1397-1406.\par
\noindent
Barrow, J.D., Tipler, F.J. (1988), The Anthropic Cosmological
Principle,\par 
New York: Oxford University Press.\par
\noindent
Bernstein, J., Feinberg, G., Eds. (1986), Cosmological
Constants:\par 
Papers in Modern Cosmology, New York: Columbia University
Press.\par
\noindent
Burbidge, E.M., Burbidge, G.R., Fowler, W.A., Hoyle, F. (1957),
Reviews of Modern\par 
Physics 29, 547-560.\par
\noindent
Carr, B.J., Rees, M.J. (1979), Nature, 278, 605-612.\par
\noindent
Chandrasekhar, S. (1935), Monthly Notices of the Royal
Astronomical
Society, 95, 207-225.\par
\noindent
Clayton, D.D. (1983), Principles of Stellar Evolution and
Nucleosynthesis,\par 
Chicago: The University of Chicago Press.\par
\noindent
Dirac, P.A.M. (1937), Nature, 139, 323. \par
\noindent
Dirac, P.A.M. (1938), Proceedings of the Royal Society, A165,
199-208.\par
\noindent
Donaldson, S.K. (1996), Bulletin of the American Mathematical
Society, 33, 45-70.\par
\noindent
Dyson, F.J. (1972), in Aspects of Quantum Theory (Eds. Salam,
A., Wigner, E.P.),\par Cambridge: Cambridge University
Press, 213-236.\par
\noindent
Einstein, A. (1917), Sitzungsberichte der Preussischen Akademie
der Wissenschaften Berlin\par 
142-152 (English translation in Bernstein, J., Feinberg, G.
(1986), Cosmological\par Constants, New York: Columbia
University Press).\par
\noindent
Friedmann, A. (1922), Zeitschrift f\"{u}r Physik 10, 377-386.
(English tranlsation in\par Bernstein, J., Feinberg, G. (1986),
Cosmological Constants, New York: Columbia\par University
Press).\par
\noindent
Friedmann, A. (1924), Zeitschrift f\"{u}r Physik 21, 326-332.
(English tranlation in\par Bernstein, J., Feinberg, G. (1986),
Cosmological Constants, New York: Columbia\par University
Press).\par
\noindent
Gamow, G. (1946), Physical Review 70,572-573.\par
\noindent
Geller, M.J., Huchra, J.P. (1989), Science 246, 897-903.\par
\noindent
Gloeckler, G., Geiss, J. (1996), Nature, 381, 210-212.\par
\noindent
Guth, A.H. (1981), Physical Review D23, 347-356.\par
\noindent
Haubold, H.J., Mathai, A.M. (1994), in Basic Space Science
(Eds. Haubold, H.J.,\par 
Onuora, L.I.), New York: American Institute of Physics, 102-
116.\par
\noindent
Hawking, S.W. (1974), Nature, 248, 30-31.\par
\noindent
Hoyle, F., Tayler, R.J. (1964),  Nature 203, 1108-1110.\par
\noindent
Kaku, M. (1993), Quantum Field Theory, New York: Oxford
University Press.\par
\noindent
Kolb, E.W., Turner, M.S. (1988), The Early Universe: Reprints,
Redwood City,\par California: Addison-Wesley Publishing
Company.\par
\noindent
Lema\^{\i}tre, G. (1927), Annales de la Soci\'{e}t\'{e}
scientifique de Bruxelles\par A47, 49 (English translation in
Bernstein, J., Feinberg, G. (1986), Cosmological\par Constants,
New York: Columbia University Press).\par
\noindent
Manin, Yu.I. (1981), Mathematics and Physics, Boston:
Birkh\"{a}user.\par
\noindent
North, J.D. (1965), The Measure of the Universe: A History of
Modern Cosmology,\par 
Oxford: Clarendon Press.\par
\noindent
North, J.D. (1995), The Norton History of Astronomy and
Cosmology, New York:\par 
W.W. Norton and Company.\par
\noindent
Pais, A. (1986), Inward Bound: Of Matter and Forces in the
Physical World, New York:\par 
Oxford University Press.\par
\noindent
Peebles, P.J.E. (1966),  The Astrophysical Journal
146, 542-552.\par
\noindent
Penzias, A.A., Wilson, R.W. (1965), The Astrophysical Journal
142, 419-421.\par
\noindent
Rees, M.J., Ostriker, J.P. (1977), Monthly Notices of the Royal
Astronomical Society, 179, 541-559.\par
\noindent
Schmid, R. (1992), SIAM Review, 34, 406-425.\par
\noindent
Sciama, D.W. (1953), Monthly Notices of the Royal Astronomical
Society, 113, 34-42.\par
\noindent
Shaver, P.A., Wall, J.V., Kellerman, K.I., Jackson, C.A.,
Hawkins, M.R.S.\par 
(1996), Nature, 384,439-441.\par
\noindent
Silk, J.I. (1977) The Astrophysical Journal 211,638-648.\par
\noindent
Smoot, G., Davidson, K. (1993), Wrinkles in Time: The Imprint
of Creation,\par 
London: Abacus.\par
\noindent
Tytler, D., Fan, X.-M., Burles, S. (1996), Nature,
381, 207-209.\par
\noindent
Wagoner, R.V., Fowler, W.A., Hoyle, F. (1967),  The
Astrophysical
Journal 148, 3-49.\par
\noindent
Weisskopf, V.F. (1975), Science, 187, 605-612.\par
\noindent
White, S.D., Rees, M.J. (1978), Monthly Notices of the Royal
Astronomical Society 183, 341.\par
\noindent
Wilczek, F. (1996), Nature, 380, 19-20.\par
\noindent
Whittaker, E. (1951 and 1953/1989), A History of the Theories
of Aether and
Electricity,\par 
New York: Dover Publications.
\noindent
{\large\bf Further Reading List}\par
\noindent
Beer, A. (1898/1961), A Short History of Astronomy: From the
Earliest Times\par 
Through the Nineteenth Century, New York:
Dover Publications.\\
Celnikier, L.M. (1989), Basics of Cosmic Structures,
Gif-sur-Yvette Cedex, France:\par Editions Frontieres.\\
Chandrasekhar, S. (1990), Truth and Beauty, Chicago: The
University of Chicago Press.\\
Collins II, G.W. (1989), The Fundamentals of Stellar
Astrophysics,  New York:\par 
W.H. Freeman and Company.\\
Combes, F., Boisse, P., Mazure, A., Blanchard, A. (1995),
Galaxies and Cosmology,\par Berlin: Springer.\\
\noindent
Danby, J.M.A., Kouzes, R., Whitney, C. (1995), Astrophysics
Simulations (The Consortium\par for Upper-Level Physics
Software), New York: John Wiley and Sons.\\
Kolb, E.W., Turner, M.S. (1990), The Early Universe, Redwood
City, California:\par Addison-Wesley Publishing Company.\\
Lang, K.R. (1980), Astrophysical Formulae: A Compendium for\par 
the Physicist and Astrophysicist, New York: Springer-Verlag.\\
Lawrie, I.D. (1994), A Unified Grand Tour of Theoretical
Physics, Bristol: Institute\par 
of Physics Publishing.\\
Misner, C.W., Thorne, K.S., Wheeler, J.A. (1973), Gravitation,
New York:\par 
W.H. Freeman and Company.\\
\noindent
Padmanabhan, T. (1993), Structure Formation in the Universe,
Cambridge:\par 
Cambridge University Press.\\
Pagels, H.R. (1982), the Cosmic Code, New York: Bantam Books.\\
Pagels, H.R. (1985), Perfect Symmetry: The Search of the
Beginning\par 
of Time, New York: Bantam Books.\\
Pagels, H.R. (1988), the Dreams of Reason, New York: Bantam
Books\\
Peebles, P.J.E. (1980), The Large-Scale Structure of the
Universe, Princeton, New Jersey:\par 
Princeton University Press.\\
\noindent
Peebles, P.J.E. (1993), Principles of Physical Cosmology,
Princeton, New Jersey:\par Princeton University Press.\\
\noindent
Rowan-Robinson, M. (1996), Cosmology, Oxford: Clarendon Press.\\
Rozental, I.L. (1988), Big Bang - Big Bounce: How Particles and
Fields Drive Cosmic\par Evolution, Berlin: Springer.\\
Stevens, C.F. (1995), The Six Core Theories of Modern Physics,
Cambridge,\par 
Massachusetts: The MIT Press.\\
Thorne, K.S. (1994), Black Holes and Time Warps, New York: W.W.
Norton and Company.\\
Trainor, L.E.H., Wise, M.B. (1981), From Physical Concept to
Mathematical Structure:\par 
An Introduction to Theoretical Physics, Toronto: University of
Toronto Press.\\
Weinberg, S. (1972), Gravitation and Cosmology: Principles and
Applications of the \par
General Theory of Relativity, New York: John Wiley and Sons.\\
\clearpage

\noindent
Figure Captions\par
\bigskip
\noindent
Fig.1. It must be related to geometry and it may have been a
window to the stars. Stonehenge in Wiltshire, England, still the
focus of hot debates.\par
\medskip
\noindent
Fig.2. Shiva is one of the two pricipal Hindu gods (the other
being Vishnu); his most celebrated appearance being the one as
Nataraja, the King of Dancers. Indian artists have represented
Shiva's cosmic dance in magnificent brone sculptures of dancing
figures with four arms whose superbly balanced and yet dynamic
gestures express the rhythm and unity of life. The upper right
hand of the god holds a drum to symbolize the primal sound of
creation, the upper left bears a tongue of flame, the element of
destruction. The balance of the two hands represents the dynamic
balance of creation and destruction in the univese, accenturated
further by the Dancer's calm and detached face in the centre of
the two hands, in which the polarity of creation and destruction
is dissolved and transcended. The second right hand is raised in
the sign of 'do not fear', symbolizing maintenance, protection
and peace, while the remaining left hand points down to the
uplifted foot which symbolizes release from the spell of maya.
The god is pictured as dancing on the body of a demon, the symbol
of human ignorance which has to be conquered befor liberation can
be attained.\par
\medskip
\noindent
Fig.3. Dante Alighieri's (1265-1321) scheme of the Universe in
illustration from ``Paradies'' in ``The Divine comedy'' extends
Aristotelian cosmology in a modern way. Dante traverses the
material world from the icy core of Earth, the abode of Lucifer,
to the Mount of Purgatory. He continues through the nine heavenly
spheres, each sphere larger and more rapidly turning than the
last, until he reaches the Primum Mobile, the ninth and largest
sphere and the boundary of space. His goal was to see the
Empyrean, the abode of God.\par
\medskip
\noindent
Fig.4. The first diagram to illustrate the proposal that the
Universe is infinite. From the edition by Thomas Digges of his
father's A Prognostication everlastinge..., published in 1576 in
London, eight years before its publication by Giordano Bruno to
whom the idea is often credited (By permission of The Royal
Society).\par
\medskip
\noindent
Fig.5. The continuum of size and organizational level appears
throughout the range of structural elements, from the fundamental
particles to the highest-level systems of matter. Particles such
as the quarks are known to be bound by strong forces. Protons and
neutrons are bound within the atomic nucleus. The outer shell of
atoms is bound to the nucleus by electromagnetic forces. A
molecule may be thought of either as a structure build of atoms
bound together by chemical forces or as a structure in which two
or more nuclei are maintained in some definite geometrical
configuration by attractive forces from a surrounding cloud of
negative electrons. The evolution of the universe created a
hierarchy of structural elements. \par
\medskip 
\noindent
Fig.6. The Big Bang: From the origin of the universe to its
present epoch
(Smoot and  Davidson, 1995).\par
\medskip
\noindent
Fig.7. Cosmology has been actively investigating the consequences
of a new extension of the theory of matter within the evolution
of the universe, in which the electricity, magnetism, weak force,
strong force, and gravity are all unified at sufficiently high
temperatures.\par
\medskip
\noindent
Fig.8. Three full-sky maps made by the COBE satellite DMR
instrument show (Smoot and Davidson, 1995).\\
Top: The dipole anisotropy caused by the Earth's motion relative
to the cosmic background radiation (hotter in the direction we
are going, cooler in the direction we are leaving).\\
Center: The dipole-removed sky showing the emission from the
plane of the galaxy - the horizontal red strip and the
large-scale ripples in space-time.\\
Bottom: A map of the wrinkles in time.\par
\medskip
\noindent
Fig.9. The Big Bang, with inflation producing space and the
ripples in space-time mapped by COBE, and eventually evolving to
become stars and galaxies and clusters of them (Smoot and
Davidson, 1995).\par
\medskip
\noindent
Fig.10. A map of the nearby universe toward the north and south
poles of the Milky Way. Each of the 9325 points in the image
represents a galaxy similar to the Milky Way. The arcs which form
the boundaries  of the two wedge-like portions of the map are all
at a distance of about 400 million light years from the Sun
(Earth). The dark regions to the east and west are obscured by
the plane of the Milky Way.
The map shows that galaxies are arranged in patterns on an
enormous scale. The Great Wall, a sheet containing thousands of
galaxies, stretches nearly horizontally across the entire
northern portion of the survey. A similar Southern Wall runs
diagonally across the southern region. These walls delineate
enormous dark voids where there are few if any galaxies. The
voids are often 150 million light years in diameter. The patterns
in the north and south are similar. These large patterns are a
tough challenge for our attempts to model the development of
structure in the universe. The curved boundaries are lines of
constant declination (Galactic latitude): in the north they are
at 8.5$^o$ and 44.5$^o$ (the wedge subtends an angle of 36$^o$ in
the narrow direction). It runs from 8 to 17 hours in right
ascension (longitude) or of order 120$^o$. In the south the
declination runs from 0$^o$ to -40$^o$ and the right ascension
runs from 20.8 to 4 hours.\\
By permission of Margaret J. Geller, John P. Huchra, Luis A. N.
da Costa, and Emilio E. Falco, Smithsonian Astrophysical
Observatory \copyright 1994\par
\clearpage
Table 1. Fundamental fermions divided into two groups: leptons
and quarks (Barnett et al., 1996).\par
\bigskip
\begin{tabular}{|c|c|c|c|c|c|c|}\hline
\multicolumn{7}{|c|}{\bf LEPTONS  spin=1/2} \\ \hline
\multicolumn{1}{|c|}{\bf Electric} & \multicolumn{1}{|c|}{\bf
Flavor} & \multicolumn{1}{|c|}{\bf Mass}  &
\multicolumn{1}{|c|}{\bf Flavor} & \multicolumn{1}{|c|}{\bf Mass}
& \multicolumn{1}{|c|}{\bf Flavor} & \multicolumn{1}{|c|}{\bf
Mass} \\ 
{\bf charge} & & $GeV/c^2$ &  & $GeV/c^2$ & &$GeV/c^2$ \\ \hline
& $\nu_e$ & & $\nu_\mu$ & & $\nu_\tau$ & \\
0 & electron & $<1\times 10^{-8}$ & muon & $1.7 \times 10^{-4}$ &
tau & $<2.4 \times 10^{-3}$\\
& neutrino & & neutrino & & neutrino & \\ \hline
& e & & $\mu$ & & $\tau$ & \\ 
-1 & electron & $5.1\times 10^{-4}$ & muon & 0.106 & tau &
1.777\\ \hline
\multicolumn{7}{|c|}{\bf QUARKS spin=1/2} \\ \hline
\multicolumn{1}{|c|}{\bf Electric} & \multicolumn{1}{|c|}{\bf
Flavor} & \multicolumn{1}{|c|}{\bf Approx. mass} &
\multicolumn{1}{|c|}{\bf Flavor} & \multicolumn{1}{|c|}{\bf
Approx. mass} & \multicolumn{1}{|c|}{\bf Flavor} &
\multicolumn{1}{|c|}{\bf Approx. mass}\\
{\bf charge} &  & $GeV/c^2$ &  & $GeV/c^2$ &  & $GeV/c^2$\\
\hline 
& {\bf u} & & {\bf c} & & {\bf t} &  \\
2/3 & up & $(2-8)\times 10^{-3}$ & charm & 1-1.6 & top & 180 \\
\hline
& {\bf d} & & {\bf s} & & {\bf b} & \\
-1/3 & down & $(5-15)\times 10^{-3}$ & strange & 0.1-0.3 & bottom
& 4.1-4.5\\ \hline
\end{tabular}
\clearpage
\begin{tabular}{|c||c|c|c|c|}\hline
\multicolumn{5}{|c|}{\bf BOSONS}\\ 
\multicolumn{5}{|c|}{spin = 0,1,2,...}\\ \hline \hline
{\bf Unified} & \hspace{2cm} & \hspace{2cm} & \hspace{2cm} &
\hspace{2cm}\\
{\bf Elektroweak} & $\gamma$ &  $W^-$ &  $W^+$ &  $Z^0$ \\
spin = 1 & photon & & & \\ \hline
{\bf Electric charge} & 0 & -1 & +1 & 0 \\ \hline
{\bf Mass} $Gev/c^2$ & 0 & 80 & 80 & 91 \\ \hline\hline
{\bf Strong or Color} & g & & & \\
spin = 1 & gluon & & & \\ \hline
{\bf electric charge} & 0 & & &  \\ \hline
{\bf Mass} $GeV/c^2$ & 0 & & & \\ \hline \hline
\end{tabular}\par
\bigskip
Table 2. Properties of the force carrriers (Barnett et al.,
1996).\par
\clearpage
\begin{tabular}{llll}\hline
\multicolumn{1}{l}{Interaction} & \multicolumn{2}{l}{Coupling
constant} &\multicolumn{1}{l}{Interaction}\\ \cline{2-3}
& Analytic  & Numerical & Radius\\
& expression & value at & cm\\
& & $m=m_p$ & \\ \hline
Gravitational & $Gm^2/\hbar c$ & $\sim 10^{-38}$ & $\infty$\\
Weak & $g_Fm^2 c/\hbar^3$ & $10^{-5}$ & $10^{-17}$\\
Electromagnetic & $e^2/\hbar c$ & $1/137$ & $\infty$\\
Strong & $a/[ln(m/m_p)]$ & $\approx 1$ & $10^{-13}$\\
& $m>>m_p$\\ \hline
\end{tabular}\par
\bigskip
Table 3. Properties of the four fundamental interactions
\clearpage
\bigskip
\begin{tabular}{lcc}\\ \hline\hline
Object & Mass & Radius \\
& $g$ & $cm$ \\ \hline
jupiter & $2\times 10^{30}$ & $6\times 10^9$ \\
sun & $2\times 10^{33}$ & $7\times 10^{10}$ \\
red giant & $(2-6)\times 10^{34}$ & $10^{14}$ \\
white dwarf & $2\times 10^{33}$ & $10^8$ \\
neutron star & $3\times 10^{33}$ & $10^6$ \\
glob. cluster & $1.2\times 10^{39}$ & $1.5\times 10^{20}$ \\
open cluster & $5\times 10^{35}$ & $3\times 10^{19}$\\
spiral & $2\times (10^{44}-10^{45})$ & $(6-15)\times 10^{22}$ \\
elliptical & $2\times(10^{43}-10^{45})$ & $(1.5-3)\times 10^{23}$
\\
group & $4\times 10^{46}$ & $3\times 10^{24}$ \\
cluster & $2\times 10^{48}$ & $1.2\times 10^{25}$ \\
\hline \hline
\end{tabular}\par
\bigskip
Table 4. Structural elements of the universe

\end{document}